\documentclass{article}
\pdfoutput=1
\usepackage{amsmath,amsfonts,amsthm,graphicx}
\usepackage{amssymb}
\usepackage{epsfig,graphicx,caption,subcaption, color}
\usepackage{mathrsfs}
\usepackage[active]{srcltx}
\usepackage{url}
\usepackage{float}
\usepackage{authblk}
\usepackage{fixme}
\usepackage[colorlinks,pdftex]{hyperref}
\usepackage{listings}
\usepackage{fixltx2e}

\newcommand{\R}{\mathbb{R}}

\renewcommand{\H}{\mathbb{H}}

\newcommand{\spc}[1]{\mathbb{#1}}

\def\d{{\rm d}}

\def\>{\rangle}
\def\<{\langle}

\newcommand{\st}[1]{\mathbf{#1}}

\DeclareMathOperator{\gap}{gap}
\DeclareMathOperator{\spp}{sp}

\DeclareMathOperator{\abs}{abs}
\DeclareMathOperator{\nneg}{neg}
\DeclareMathOperator{\poly}{poly}
\DeclareMathOperator{\av}{av}

\newtheorem{theo}{Theorem}

\newtheorem{lemma}[theo]{Lemma}

\DeclareMathOperator*{\arccosh}{arccosh}

\begin{document}
\renewcommand{\b}{\mathbf}
\newcommand{\p}{\partial}

\title{The performance of the quantum adiabatic algorithm on spike Hamiltonians}

\author{Linghang Kong}
\affil{Institute for Interdisciplinary Information Sciences, Tsinghua University,\\ Beijing 100084, P.R.China}
\author{Elizabeth Crosson}
\affil{Institute for Quantum Information and Matter, California Institute of Technology, \\ Pasadena, CA 91125, USA}

\date{}

\maketitle
\begin{abstract}
Perturbed Hamming weight problems serve as examples of optimization instances for which the adiabatic algorithm provably out performs classical simulated annealing.  In this work we study the efficiency of the adiabatic algorithm for solving the ``the Hamming weight with a spike'' problem by using several methods to compute the scaling of the spectral gap at the critical point, which apply for various ranges of the height and width of the barrier.
Our main result is a rigorous polynomial lower bound on the minimum spectral gap for the adiabatic evolution when the bit-symmetric cost function has a thin but polynomially high barrier.   This is accomplished by the use of a variational argument with an improved ansatz for the ground state, along with a comparison to the spectrum of the system when no spike term is present.   We also give a more detailed treatment of the spin coherent path-integral instanton method which was used by Farhi, Goldstone, and Gutmann in arXiv:quant-ph/0201031, and consider its applicability for estimating the gap for different scalings of barrier height and width. We adapt the discrete WKB method for an abruptly changing potential, and apply it to the construction of approximate wave functions which can be used to estimate the gap. Finally, the improved ansatz for the ground state leads to a method for predicting the location of avoided crossings in the excited states of the energy spectrum of the thin spike Hamiltonian, and we use a recursion relation to determine the ordering of some of these avoided crossings, which may be a useful step towards understanding the diabatic cascade phenomenon which occurs in spike Hamiltonians.
\end{abstract}

\section{Introduction}
Since the introduction of quantum annealing (QA)~\cite{Nishimori:98a, FGG+00}  there has been an effort to understand when the method can outperform competing classical optimization algorithms.   QA has been frequently compared with classical simulated annealing~\cite{KGV83, Ce85} because both algorithms are inspired by the physics of minimizing energy, and both can be applied without making any assumptions about the structure of the problem.   Simulated annealing (SA) has been contrasted with QA in several benchmarking studies~\cite{Martonak-2002, Santoro-2005, Boixo13, Liu14}, and in some cases analytical comparisons between the methods have been made as well~\cite{Sei09, Nishi14}.   

An example for which QA exponentially outperforms SA is a bit-symmetric problem called the ``the Hamming weight with a spike''~\cite{FGG02}, which consists of a linear potential and a large barrier which creates a false minimum.  For an $n$-bit string with Hamming weight $w$, the spike energy function is defined to be
\begin{equation}\label{eq:originalSpike}
 h(w) = 
 \begin{cases}
  w & w \not= n/4 \\
  n & w = n / 4
 \end{cases}.
\end{equation}
As usual, the proof that QA solves this problem efficiently will use the adiabatic theorem together with a lower bound on the scaling of the spectral gap with the system size.  Numerical evidence of the scaling was given in \cite{FGG02}, and a rigorous \textbf{variational-comparison argument}~\cite{Re04} was used to show that the system has a constant gap when the spike height scales as $\mathcal{O}(n^{1/2})$.  The main result of this work is an improvement of the variational-comparison argument that demonstrates an inverse polynomial lower bound on the scaling of the gap when the height of the spike is $n^{\alpha}$, for any constant $\alpha > 0$.  

We also consider a modification of the cost function \eqref{eq:originalSpike} that includes a spike with width $n^\beta$, and although we do not rigorously bound the gap in this case we obtain estimates by a variety of approximation methods:
\begin{itemize}
\item The \textbf{spin coherent path integral instanton method} has been applied to bit-symmetric QA Hamiltonians with cubic cost functions~\cite{FGG02}, to modified adiabatic paths~\cite{FGG02b}, and the importance of the spin coherent state effective potential to spike Hamiltonians was also recently emphasized by~\cite{MAL15-2}, which appeared contemporaneously with the present work.   However, many mathematical details of the method have not appeared in previous works on adiabatic computing.  We address this in Appendix~\ref{app:instanton} with a derivation of several formulas which are used to extract quantitative results from the spin coherent instanton method for bit-symmetric QA Hamiltonians, including equations (25) and (26) in \cite{FGG02}, which in that work were stated without explanation.  We use these results to obtain the predictions of the method for the spike of height $n^\alpha$ and width $n^\beta$.  

\item The \textbf{discrete WKB method} has previously been applied to spin tunneling in symmetric spin systems~\cite{Ga00}, and in the context of QA Hamiltonians with bit-symmetries~\cite{Boixo-2014}.   Here we show how to apply the discrete WKB method to make a quantitative estimate for the spectral gap of bit-symmetric QA Hamiltonians such as the spike.  Our application also requires a modification of the discrete WKB method for the case of an abruptly varying potential, and we adapt previously known techniques for continuous systems~\cite{Am14} to the discrete setting. 

\item The scaling of the gap for a spike with equal height and width was determined numerically in~\cite{BD15}, and we apply similar methods to outline the region in the $(\alpha,\beta)$ plane for which the gap decreases superpolynomially.  The numerical results support the conclusions obtained from the spin coherent instanton method and from the discrete WKB method.  
\end{itemize}

Finally, we consider the excited states of the width 1 spike Hamiltonian and observe a method for predicting the locations of avoided crossings between excited states.  This observation builds on an intuition gained by analyzing the ground state and first excited state: the avoided crossings happen when nodes in the eigenstates of the spikeless system cross the location of the spike.  We prove a theorem on the ordering of these nodal crossings, which may be useful in understanding the diabatic cascade phenomenon~\cite{MAL15}. 

The rest of this paper is structured as follows. Section~\ref{sec:preliminary} defines the generalized form of the spike Hamiltonian and details the simplifications that come from viewing the problem in the symmetric subspace.  Section ~\ref{sec:width1} contains our main result, a rigorous lower bound for the gap when the height of the spike is $n^{\alpha}$ and the width is 1. We also demonstrate an upper bound on the gap for the width 1 spike when $\alpha > 1$, which rigorously shows the existence of a critical point in the limit of large system size.  In Section~\ref{sec:general}, we consider a spike with a general polynomial width, and we apply the spin coherent instanton method and the discrete WKB method to approximate the spectral gap of this system. Finally, in Section~\ref{sec:diabatic} we discuss the locations of the avoided crossings in the higher eigenstates of the width 1 spike.  

\section{Preliminaries}
\label{sec:preliminary}
\subsection{Definitions}
We consider a natural generalization of the cost function \eqref{eq:originalSpike},
\[
 h(w) = 
 \begin{cases}
  w + \frac{3}{4}n^\alpha & n/4 - n^\beta/2 < w < n/4 + n^\beta/2 \\
  w & \textrm{o.w.} \\
 \end{cases},
\]

with a spike centered on $n/4$ of width $n^{\beta}$ and height $n^{\alpha}$ (we assume $n$ is a multiple of 4).  By the same argument in \cite{FGG02} it follows that classical simulated annealing requires
superpolynomial time to find the true minimum of this function for any $\alpha, \beta > 0$.

Reviewing the application of the adiabatic algorithm to this cost function, the Hamiltonian whose ground state encodes the solution is given by
\begin{equation}
 \mathcal H_p = \sum_{z \in \{0, 1\}^n} h(|z|) |z\>\<z|, \label{eq:hp}
\end{equation}
where $|z|$ is the Hamming weight of $z$.
We follow the convention in \cite{FGG02} and define the initial Hamiltonian as
\begin{equation}
 \mathcal H_0 = \sum_{k = 1}^n \frac{1}{2}\left(1-\sigma_x^{(k)}\right), \label{eq:h0}
\end{equation}
where $\sigma_x^{(k)}$ is the Pauli $x$ operator acting on the $k$-th qubit.
The ground state of $\mathcal H_p$ is prepared by initializing the system in the ground state of $\mathcal{H}_0$, and linearly interpolating between $\mathcal H_0$ and $\mathcal H_p$, 
\begin{equation}
 \mathcal H(s) = (1-s) \mathcal H_0 + s \mathcal H_p, \label{eq:hamiltonian}
\end{equation}
where $0 \leq s \leq 1$ is called the adiabatic parameter.  The system is initialized with $s = 0$ to the ground state of $\mathcal H_0$, which is $[(|0\> + |1\>)/\sqrt 2]^{\otimes n}$, and the adiabatic parameter is slowly increased so that the system remains close to the ground state of the instantaneous Hamiltonian throughout the evolution~\cite{BF28}.  The total running time scales polynomially with the inverse of the minimum spectral gap~\cite{Re04,Jansen07}.

\subsection{Simplification in symmetric subspace}
The Hamiltonian~\eqref{eq:hamiltonian} and the initial state $[(|0\> + |1\>)/\sqrt 2]^{\otimes n}$ are invariant under permutation of the qubits, so the state at all future times will belong to the symmetric subspace as well.  A basis for this subspace is $\{|k\>\}_{k=0}^n$, where
\[
 |k\> = \frac{1}{\sqrt{\binom{n}{k}}}\sum_{z: |z| = k}|z\>.
\]
Note that this is also the subspace with the largest total angular momentum of the $n$ qubits, when each qubit is envisioned as a spin-1/2 particle. Thus we can regard the system as a spin $J$ particle with $J = n/2$, and for this system $X = \sum_{k=1}^n \sigma_x^{(k)}/2$ is the angular momentum operator in the $x$-direction. The $z$-direction angular momentum operator $Z$ is
\[
 Z = \sum_{k=0}^n \left(J - k\right)|k\>\<k|.
\]
When restricted to the symmetric subspace, the Hamiltonian simplifies to
\begin{align}
 \mathcal H(s) &= \frac{1-s}{2} \sum_{k = 1}^n \left(1 - \sigma_x^{(k)}\right) + \frac{s}{2} \sum_{k = 1}^n \left(1 - \sigma_z^{(k)}\right) + s H_{\spp}\nonumber \\
  &= J - (1-s)X - sZ + s H_{\spp}, \label{eq:symmetric-h}
\end{align}
where $H_{sp}$ is a spike term with height $n^\alpha$ on the set $R_\beta =\{k :n/4 - n^\beta/2 \leq k \leq n/4 + n^\beta/2\}$,
\[
 H_{sp} = \frac{3}{4}n^\alpha\sum_{k\in R_\beta}|k\>\<k|.
\]
Note that the constant term $J$ does not affect the gap. Consider
\begin{equation}
  H(s) = - \sin\theta X - \cos\theta Z + \cos\theta H_{\spp} \label{eq:Hamiltonian}
\end{equation}

where $\theta$ is a function of $s$ satisfying
\[
 \sin\theta = \frac{1-s}{\sqrt{s^2 + (1-s)^2}}, \quad \cos\theta = \frac{s}{\sqrt{s^2 + (1-s)^2}}.
\]
The gap of $\mathcal H(s)$ will be $\sqrt{s^2 + (1 - s)^2}$ times that of $ H(s)$. Since we are only interested the scaling of the gap while $\sqrt 2/2 \le \sqrt{s^2 + (1 - s)^2} \le 1$ is a constant factor, we will only consider $ H(s)$ in the rest of the paper.

The ground state of the Hamiltonian without the spike term is given by
\[
 \sum_{k=0}^n \left[\<k|\left(\cos\frac{\theta}{2}|0\> + \sin\frac{\theta}{2}|1\>\right)^{\otimes n}\right]|k\> 
 = \sum_{k=0}^n \sqrt{\binom{n}{k}}\sin^k\frac{\theta}{2}\cos^{n-k}\frac{\theta}{2}|k\>.
\]

This is the square root of a Binomial distribution $\mathcal{B}(n, \sin^2(\theta/2))$.  The peak of this state coincides with the spike at $s = s^* \equiv (\sqrt 3 - 1) / 2 \approx 0.366$.  In the next section we will show that this $s^*$ is a critical point for the system with the spike, with the caveat that our upper bound on the scaling of the gap with system size only holds for  $\alpha > 1$. However, numerical evidence (see Fig.~\ref{fig:slope-alpha}) also indicates that $s^*$ is a critical point of the system for all $1/2 \leq \alpha \leq 1$.  Note that our lower bound for the scaling of the gap still applies when $\alpha < 1$, so the adiabatic run-time will be polynomial in that case as long as $s^*$ remains the location of the minimum gap as the numerics suggest.   

\section{Spike with width 1}\label{sec:width1}
This section will be devoted to the proof of our main theorem:
\begin{theo}
\label{thm:width1}
The Hamiltonian \eqref{eq:hamiltonian}, with a spike of height $n^\alpha$ and a width equal to 1, has a critical point at $s^* =(\sqrt 3 - 1) / 2$ for $\alpha > 1$.  The spectral gap of $\mathcal H(s^*)$ scales with system size as,
 \[
  {\gap}(n) =
  \begin{cases}
   \Omega(n^{1/2 - \alpha}) & \alpha > 1/2 \\
   \Omega(1) & \alpha < 1/2
  \end{cases}.
 \]

\end{theo}

\begin{figure}[ht]
 \centering
 \includegraphics[width=0.5\textwidth]{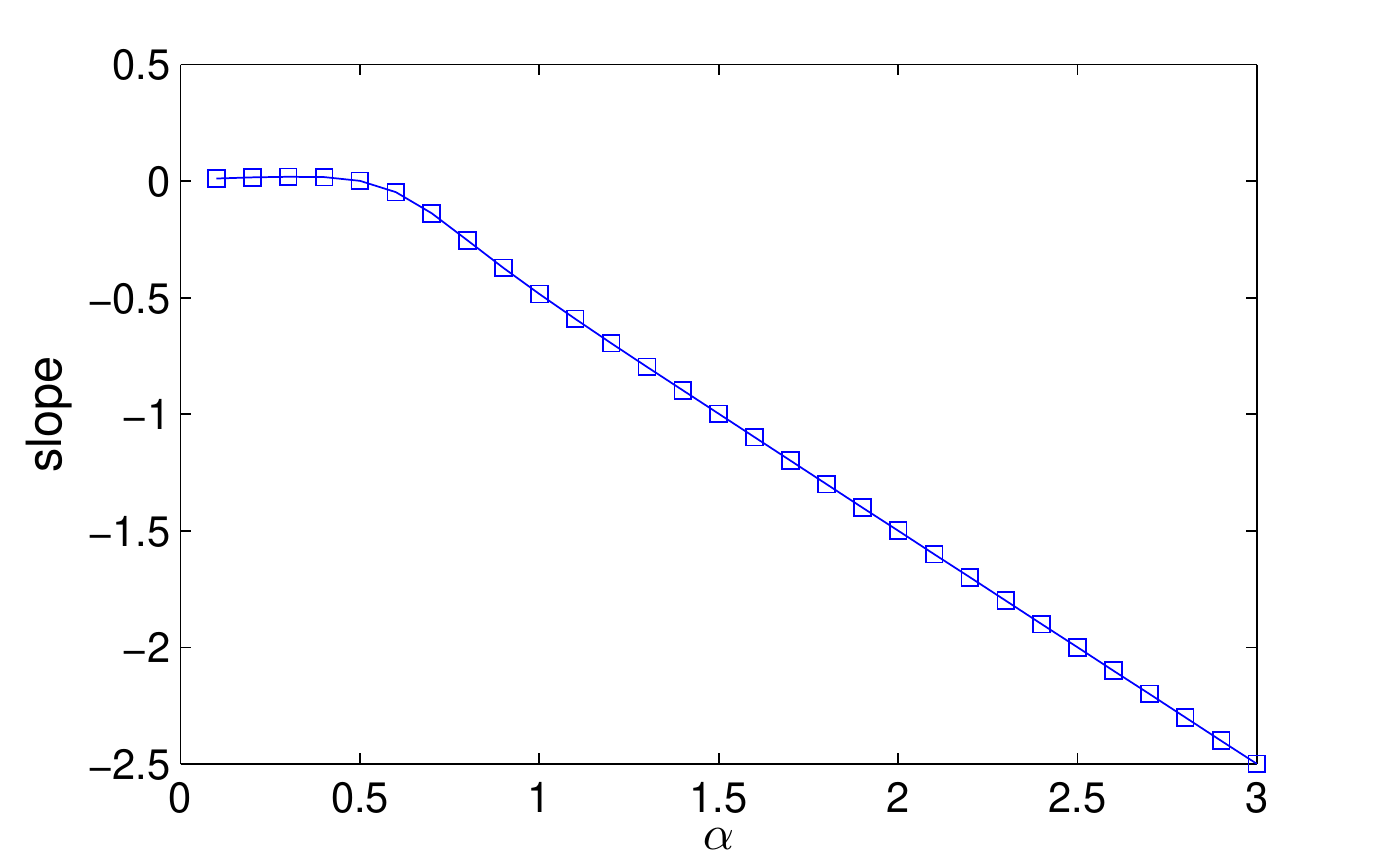}
 \caption{The slope of $\log(\gap)$ vs $\log(n)$ as a function of $\alpha$, which corresponds to the exponent $c$ in $\gap = \Theta(n^{c})$. For each value of $\alpha$, the slope is calculated from the numerical result with $n$ ranging from 500 to 860.}
 \label{fig:slope-alpha}
\end{figure}

Theorem~\ref{thm:width1} agrees with results obtained by numerical diagonalization at finite system sizes, shown in Fig.~\ref{fig:slope-alpha}. 

The rest of this section is organized as follows.  First we review Reichardt's variational-comparison method~\cite{Re04} which demonstrates that the gap is constant for $\alpha < 1/2$.  Following this we exhibit an improved trial wave function for the ground state of the spike Hamiltonian, and compute the lower bound on the spectral gap for general $\alpha >1/2$.  Finally, to obtain an upper bound on the gap when $\alpha > 1$ we use a general method for lower bounding the ground state energy of stoquastic Hamiltonians~\cite{FGG+10}, a set which includes our case \eqref{eq:hamiltonian}.

\subsection{Previous result}
The variational principle of quantum mechanics expresses the ground state energy $\bar{E}$ of a Hamiltonian $\bar{H}$ as a minimum of the expectation of $\bar{H}$ over all states in the Hilbert space $\H$,
\begin{equation}
\bar{E} = \min_{|\psi\> \in \H} \frac{\<\psi|\bar{H}|\psi\>}{\<\psi|\psi\>},
\end{equation}
from which one obtains an upper bound on the ground state energy from any trial state $|\psi\>\in \H$.  Similarly, for any subset $\spc{S}$ of the Hilbert space,
\begin{equation}
\bar{E} \leq \min_{|\psi\> \in \spc{S}} \frac{\<\psi|\bar{H}|\psi\>}{\<\psi|\psi\>}.\label{eq:variationalPrinciple}
\end{equation}
Let $E_0, E_1, \ldots$ be the energy eigenvalues of the spikeless Hamiltonian $ H(s) - \cos\theta H_{sp}$, arranged in ascending order, while $E_0', E_1', \ldots$ label the eigenvalues of the Hamiltonian with the spike $ H(s)$. It follows from the Courant-Fischer min-max theorem~\cite{MAL15-2} that $E_k' \ge E_k$ for all $k \geq 0$. Combining this with the variational principle \eqref{eq:variationalPrinciple},
\begin{align}
 \operatorname{gap}(n) & = E_1' - E_0' \\
  &\ge E_1 - E_0' \\
  &= E_1 - \min_{|\psi\> \in \spc S} \frac{\<\psi| H(s)|\psi\>}{\<\psi|\psi\>}\label{eq:gapLB}
\end{align}
where $\spc S$ is any subset of the Hilbert space.  The ground state probability distribution $|\<k|\psi_0\>|^2$ is a binomial distribution with a maximum height which is $\mathcal{O}(n^{-1/2})$, therefore using $|\psi_0\>$ as the trial wave function in \eqref{eq:gapLB} suffices to show that the gap is constant when $\alpha < 1/2$.

\subsection{Lower bound on the spectral gap}
\begin{figure}[ht]
 \centering
 \begin{subfigure}[t]{0.4\textwidth}
  \includegraphics[width=\textwidth]{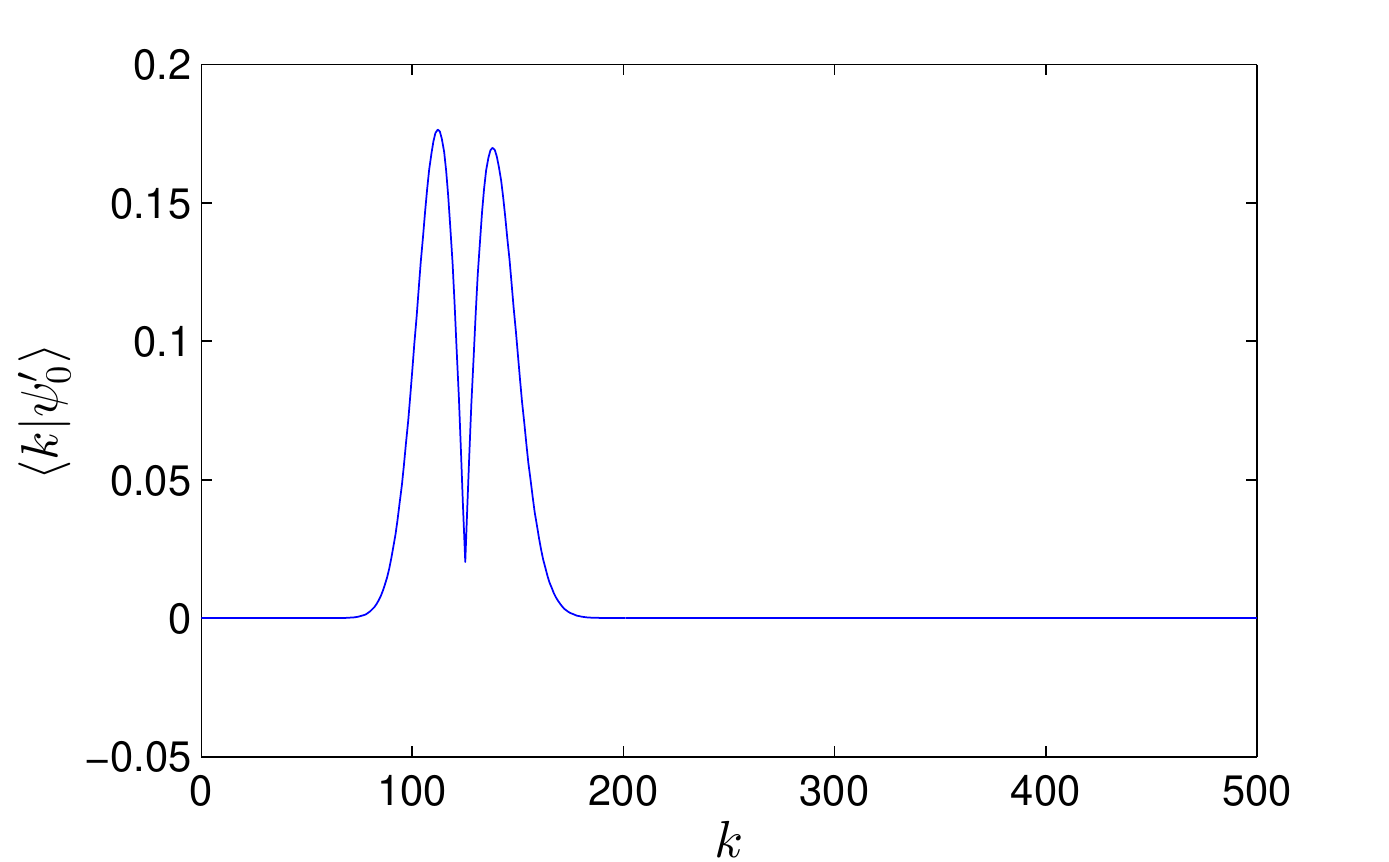}
  \caption{}
  \label{fig:g-spiked}
 \end{subfigure}
 \begin{subfigure}[t]{0.4\textwidth}
  \includegraphics[width=\textwidth]{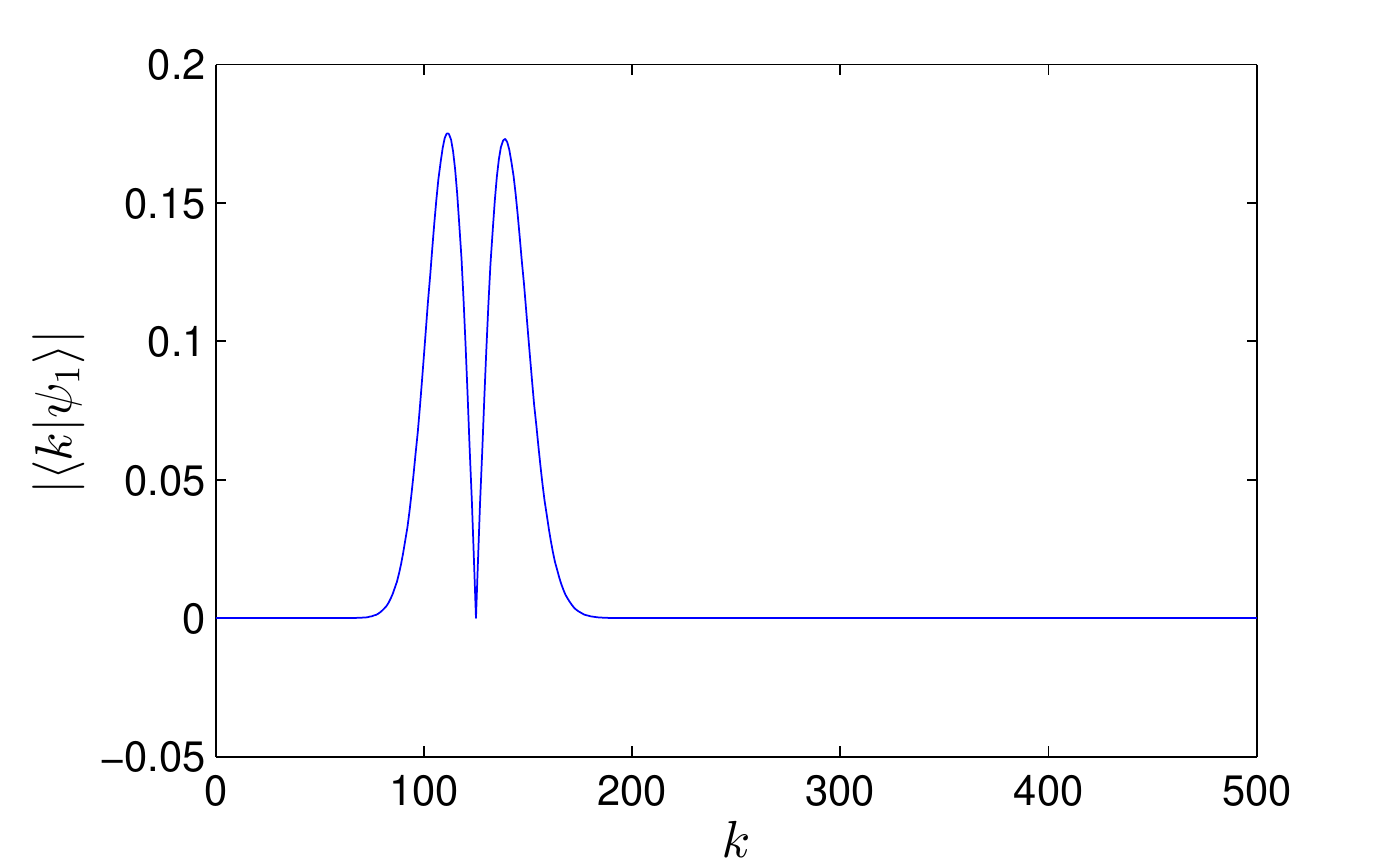}
  \caption{}
  \label{fig:abs}
 \end{subfigure}
 \caption{Fig.~\ref{fig:g-spiked} shows the ground state wave function of the spike Hamiltonian in the basis $\{|k\>\}_{k=0}^n$. Fig.~\ref{fig:abs} shows the absolute value of the first
 excited state wave function of the spikeless Hamiltonian in this basis. Both figures are generated with $n = 500$ and $\alpha = 1$.}
\end{figure}

Let $H\equiv H(s^*)$ for the remainder of this section, and let $H_0$ be the spikeless Hamiltonian at $s=s^*$,
\[
  H_0 = -\frac{\sqrt 3}{2}X - \frac{1}{2} Z.
\]
When $\alpha > 1/2$, one no longer obtains a useful bound from by taking the spikeless ground state as a trial wave function as in the previous section.  To build intuition for the design of a better trial wavefunction, the ground state probability distribution of $H$ is shown in Fig.~\ref{fig:g-spiked}.

The amplitudes for this bimodal distribution bear a close resemblance to absolute value of the amplitudes of the first excited state of the spikeless system,
\begin{equation}
\<k|\psi_{\abs}\> \equiv |\<k|\psi_1\>|,\label{eq:abs1prime}
\end{equation}
as shown in Fig.~\ref{fig:abs}.  This correspondence suggests that we minimize over the set of states,
\begin{equation*}
 \spc S = \{|\psi_{\abs}\> + x |\psi_0\>| x \in \R\}.
\end{equation*}
The full details of the calculation are contained in Appendix~\ref{app:variational}, where we show that setting
\[
 x = x_0 \equiv \frac{\sqrt 3}{2} n^{1/2 - \alpha}
\]
implies that the state $|\psi\> \equiv |\psi_{\abs}\> + x |\psi_0\>$ will satisfy
\[
 E_1 - \frac{\<\psi|H|\psi\>}{\<\psi|\psi\>} = \sqrt{\frac{3}{2\pi}} n^{1/2 - \alpha} (1 + o(1)),
\]
which together with \eqref{eq:gapLB} implies $\gap(H(s^*)) = \Omega(n^{1/2 - \alpha})$.
\subsection{Proof that \textit{s*} is a critical point}
\label{subsec:critical}
To show that $\operatorname{gap}(H(s^*))$ decreases at least inverse polynomially with system size when $\alpha > 1$, we make use of the following lemma which can be found in \cite[Theorem~4]{FGG+10}.
\begin{lemma}
\label{lem:lb}
 Let $\bar H$ be a Hermitian operator which is stoquastic in the $|k\>$ basis (meaning $\<k|\bar H|k'\> \le 0$ for all $k \not= k'$). Let $\bar E_0$ be its lowest eigenvalue. Then
\begin{equation}
  \bar E_0 \ge \min_k \frac{\<k|\bar H|\phi\>}{\<k|\phi\>}\label{eq:gap-lb}
\end{equation}
for any state $|\phi\>$ such that $\<k|\phi\> > 0$ for all $k$.
\end{lemma}

With this lemma, we can prove the following theorem.
\begin{theo}
 When $\alpha > 1$, $\lim_{n \to +\infty}\operatorname{gap}(n) = 0$ at $s^* = (\sqrt 3 - 1) / 2$.
\end{theo}

\textit{Proof.} We continue with the ansatz $|\phi\> = |\psi_{\abs}\> + x |\psi_0\>$, where $x > 0$ ensures that $|\phi\>$ satisfies the conditions in Lemma~\ref{lem:lb} (note that \eqref{eq:gap-lb} is valid even when $|\phi\>$ is not normalized). We evaluate $E_1 - \frac{\<k| H|\phi\>}{\<k|\phi\>}$ for the following two cases,
 \begin{enumerate}
  \item
  $k \not= n / 4$. In this case we have
  \[
   \<k| H|\psi_{\abs}\> = E_1\<k|\psi_{\abs}\>,
  \]
  so
  \[
   E_1 - \frac{\<k| H|\phi\>}{\<k|\phi\>} = E_1 - \frac{E_1\<k|\psi_{\abs}\> + x E_0 \<k|\psi_0\>}{\<k|\psi_{\abs}\> + x \<k|\psi_0\>}
   = \frac{(E_1 - E_0)x}{\frac{|n-4k|}{\sqrt{3n}}+x} = \frac{x}{\frac{|n-4k|}{\sqrt{3n}}+x}
  \]
  with the expressions for $\<k|\psi_{\abs}\>$ and $\<k|\psi_0\>$ given in \eqref{eq:ampl-g} and \eqref{eq:ampl-e} taken into account.
  \item
  $k = n / 4$.  Since $\<k|\psi_{\abs}\> = 0$,
  \begin{align*}
   \<k| H|\psi_{\abs}\> = \frac{\sqrt 3}{2}\<k|X|\psi_{\abs}\>  = \frac{\sqrt{3n}}{2} \sqrt{\binom{n}{k}}\left(\frac{1}{2}\right)^k\left(\frac{\sqrt 3}{2}\right)^{n-k}.
  \end{align*}
  which implies
  \begin{align*}
   E_1 - \frac{\<k| H|\phi\>}{\<k|\phi\>} 
   &= E_1 - \frac{\<k| H|\psi_{\abs}\> + E_0 x \<k|\psi_0\> + \frac{1}{2} \cdot \frac{3}{4}n^\alpha x\<k|\psi_0\>}{x \<k|\psi_0\>}
   &= \left(1 - \frac{3}{8}n^\alpha + \frac{\sqrt{3n}}{2x}\right).
    \end{align*}

 \end{enumerate}
One can readily verify that the first excited state for the spikeless Hamiltonian $ H$ is also the first excited state for the spike Hamiltonian.  Combining this fact with Lemma~\ref{lem:lb},
 \begin{align}
  \operatorname{gap}(n) &= E_1' - E_0' \nonumber \\
  &\le E_1 - \min_k \frac{\<k| H|\phi\>}{\<k|\phi\>} \nonumber \\
  &= \max\left\{\max_{k\not= n/4}\frac{x}{\frac{|n-4k|}{\sqrt{3n}}+x},
 \left(1 - \frac{3}{8}n^\alpha + \frac{\sqrt{3n}}{2x}\right)\right\}\nonumber \\
  &= \max\left\{\frac{x}{\frac{4}{\sqrt{3n}}+x}, 1 - \frac{3}{8}n^\alpha + \frac{\sqrt{3n}}{2x}\right\}, \label{eq:maxmax}
 \end{align}
 which holds for all $x>0$. Making the choice
 \[
 x = \frac{4\sqrt{3 n}}{3 n^\alpha - 8}
 \]
 in \eqref{eq:maxmax} yields
 \[
 \gap(n) \leq \max\left\{\frac{3n}{3n + 3n^{\alpha} - 8} , 0\right \} = \Theta(n^{1- \alpha}),
 \]
 which proves that for $\alpha > 1$ the gap of $H(s^*)$ goes to zero in the limit of infinite system size.

\section{A spike with polynomial height and width}
In this section we analyze the spike Hamiltonian \eqref{eq:hamiltonian} with a spike barrier of polynomial width $n^\beta$, where $\beta > 0$.  After a brief numerical investigation in Section \ref{subsec:numerics} to guide our expectations, we first apply an instanton method in Section \ref{subsec:instanton}, which approximates the gap by a tunneling rate between local minima of an effective potential on spin coherent states.  The application of the instanton method to a Hamiltonian such as \eqref{eq:hamiltonian} appears previously in~\cite{FGG02}, but here we provide more details of the calculation and apply it to our more general parametrization of the spike.   After the instanton calculation, we find that a slightly better estimate of the spectral gap can be obtained by using a discrete WKB method to construct an approximate wave function in Section \ref{subsec:WKB}.    The WKB method has been applied previously to discrete systems~\cite{Ga00,Boixo-2014}, but here we use a modification for our sharply changing potential which has to our knowledge not been previously applied in a discrete system.   In contrast with the previous section, both of these methods depend on approximations and only yield an asymptotic estimate of the scaling of the gap, rather than rigorous upper and lower bounds.  When $\beta$ is sufficiently large the assumptions of the methods hold with higher accuracy, and in this regime the methods correctly indicate that the gap goes to zero as an inverse superpolynomial.
\label{sec:general}
\subsection{Numerical estimate of the gap scaling}\label{subsec:numerics}
Before stating our numerical results for the spike Hamiltonian, we first review the results of Brady and van Dam~\cite{BD15}.  In that work the spike term is parameterized to have both a height and width scaling as $n^\alpha$, and the reported result is that the gap is constant when $\alpha < 1/4$, decreasing polynomially in system size when $1/4 < \alpha < 1/3$, and decreasing superpolynomially when $\alpha > 1/3$.   In Section \ref{subsubsec:spikeGapWKB} we show that the discrete WKB method predicts a superpolynomially decreasing gap when $\alpha + 2 \beta > 1$, which matches the finding of Brady and van Dam that the gap decreases superpolynomially for the case $\alpha = \beta > 1/3$.

\begin{figure}[H]
 \centering
 \begin{subfigure}[t]{0.45\textwidth}
  \includegraphics[width=\textwidth]{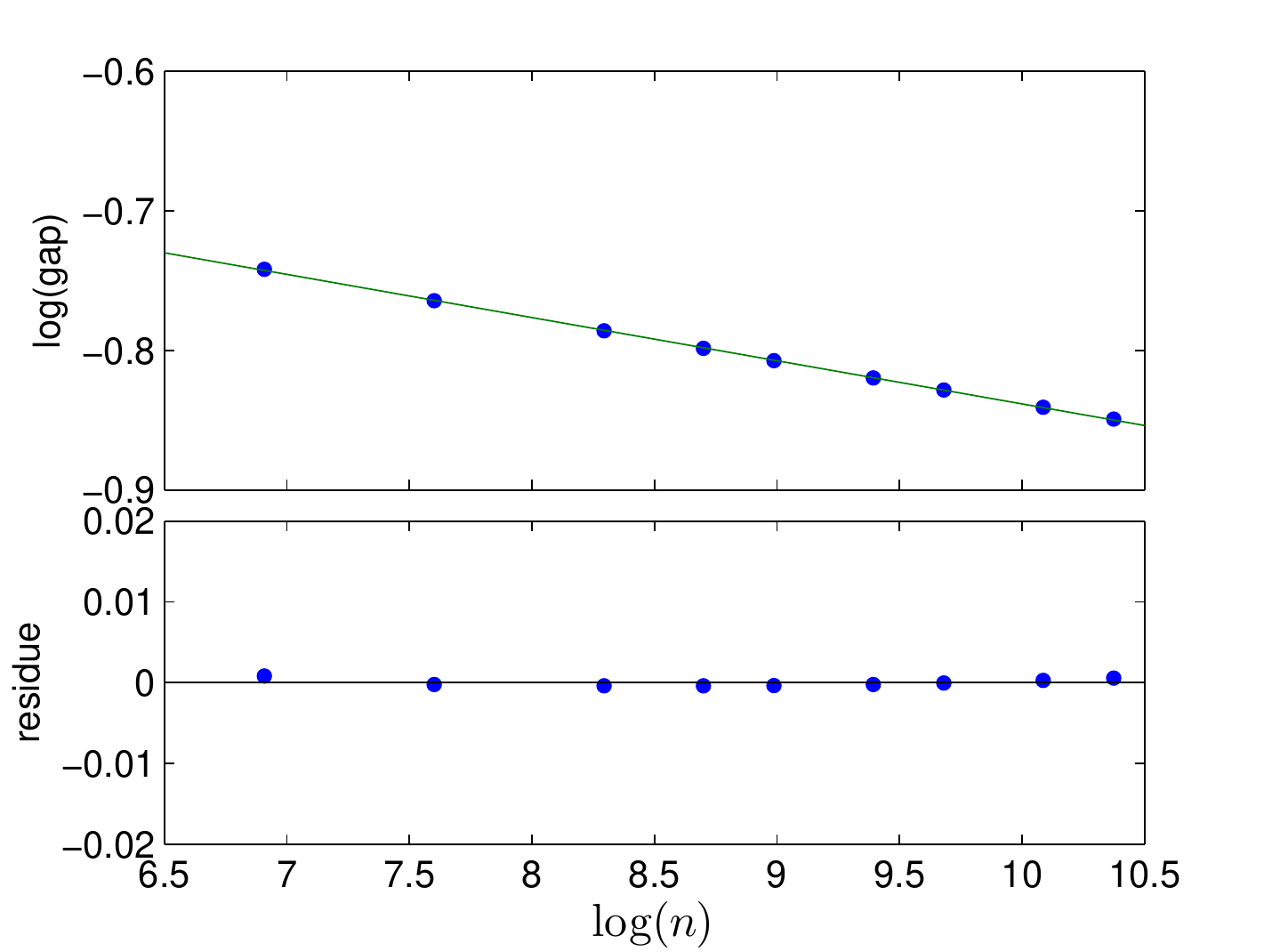}
  \caption{}
  \label{fig:subfig1}
 \end{subfigure}
 \begin{subfigure}[t]{0.45\textwidth}
  \includegraphics[width=\textwidth]{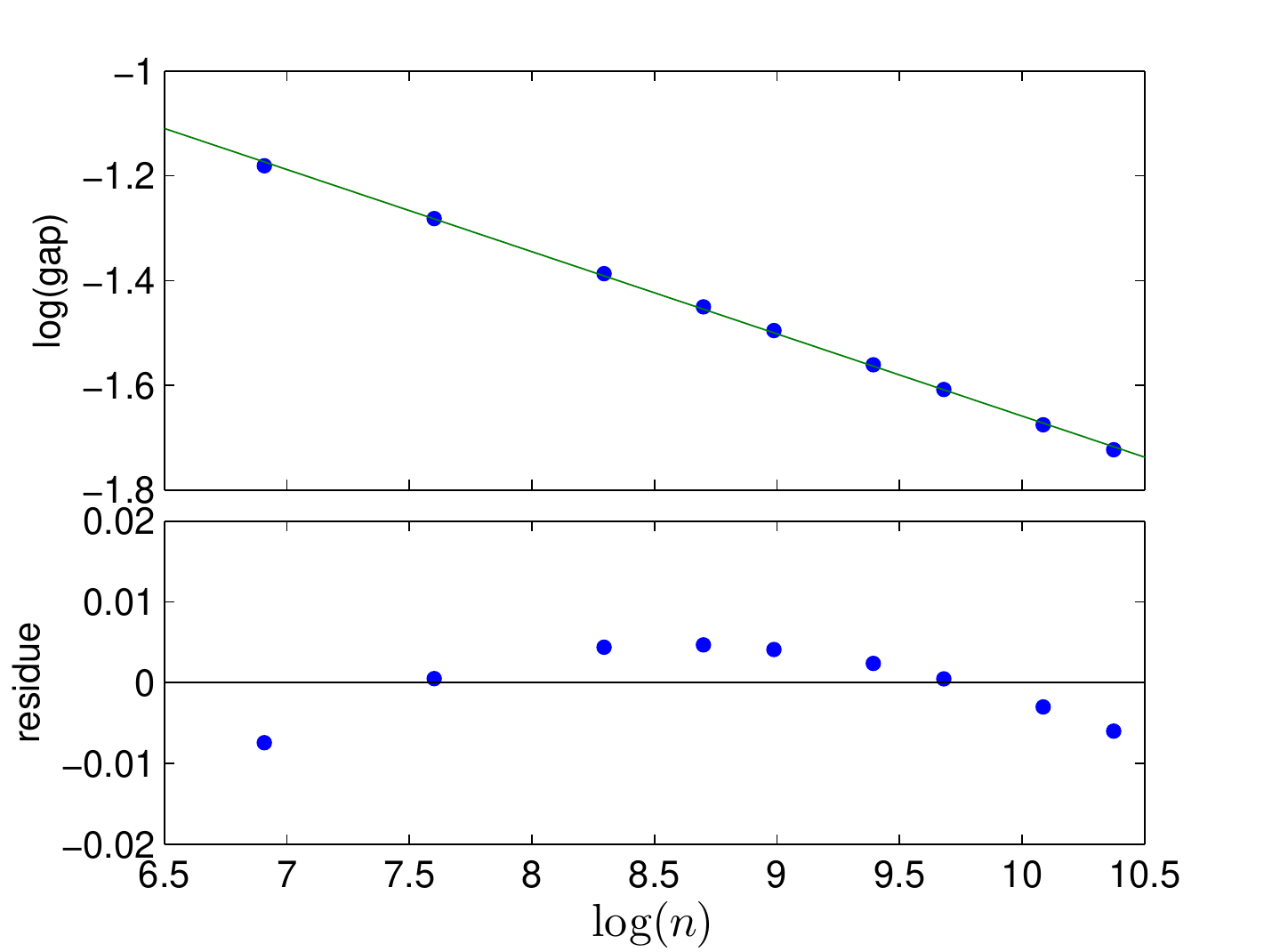}
  \caption{}
  \label{fig:subfig2}
 \end{subfigure}
 \caption{Linear fitting and the residue in $\log(\operatorname{gap}(n))$ vs $\log n$ plot.
 Fig.~\ref{fig:subfig1} describes the case $\alpha = 0.15, \beta = 0.4$. Fig.~\ref{fig:subfig2} describes the case $\alpha = 0.25, \beta = 0.4$.}
 \label{fig:wide-spike}
\end{figure} 
A challenge in this numerical analysis is that when $n$ is not large
enough, the relationship $\log(\operatorname{gap}(n))$ vs $\log n$ may appear to be a straight line when the superpolynomial function is subexponential, due to the finite system sizes ($n \lesssim 3\times 10^4$) we consider. We adopt the method used in \cite{BD15} of examining the concavity of residue of
the linear fitting in $\log(\operatorname{gap}(n))$ vs $\log n$. For example, when $\alpha = 0.15, \beta = 0.4$ the flatness of the residue in Fig.~\ref{fig:subfig1} indicates that the gap decreases polynomially, while in Fig.~\ref{fig:subfig2} the concavity of the residue indicates that the gap decreases superpolynomially.  When this analysis is applied to other choices of $\alpha$ and $\beta$ the results are qualitatively consistent with the predictions in Section~\ref{subsubsec:spikeinstantons} and Section~\ref{subsubsec:spikeGapWKB}.
\subsection{The spin coherent path integral instanton method}\label{subsec:instanton}
In quantum systems with spatial degrees of freedom, the semi-classical picture of tunneling is based on localized quantum states within the local minima of the potential.  In some cases the spectral gap of the quantum system is approximately equal to the difference between the energy of the antisymmetric and symmetric superpositions of these localized states.  The instanton method relates this energy difference to the tunneling rate through the barrier separating the minima, and computes this tunneling rate using the path integral representation of the system.   In particular, the dominant contribution to the tunneling amplitude arises from the trajectories that minimize the Euclidean action, and these trajectories are called instantons. 

For spin systems the semi-classical localized states are spin coherent states $\left\{|\mathbf{n}\rangle = |\theta,\phi\>\right\}$, defined by
\[
\left(\cos\theta S_z + \sin\theta\cos\phi S_x + \sin\theta\sin\phi S_y\right)|\theta,\phi\> = J |\theta,\phi\>,
\]
where $J = n/2$ is the total spin quantum number, and $\mathbf{n} = (\sin \theta \cos \phi, \sin \theta \sin \phi, \cos \theta)$ is a unit vector.  Consider the classical energy $\<\theta,\phi|H|\theta,\phi\>$, which leads to the following potential for $\phi = 0$,
\[U(\theta,s) = \frac{n}{2}(1-s)(1 - \sin\theta) +  s\<\theta, 0|H_p|\theta, 0\>.
\]
Spin tunneling occurs when there are degenerate local minima $|\mathbf{n}_1\>$, $|\mathbf{n}_2\>$ of this potential which are separated by an appropriately shaped barrier that the system can tunnel through.  The instanton method relates the tunneling amplitude $\mathcal{U}_{21} \equiv \<\st n_2 | e^{- T H} |\st n_1\>$ to the spectral gap of a system with a ground state and first excited state that approximately correspond to symmetric and antisymmetric combinations of states localized at $\st n_1$ and $\st n_2$ (more details of this calculation are given in Appendix \ref{app:instanton}).   Meanwhile, another independent estimate of $\mathcal{U}_{21}$ can be obtained from a path integral,
\[
\mathcal{U}_{2 1}  = \lim_{T\to\infty}\int_{\mathcal{P}} \mathcal{D}\mathbf{n(\tau)} \exp^{-  S_E[\mathbf{n}(\tau)]} \quad , \quad S_E[\mathbf{n}(\tau)] = \int_{0}^{T} \left [\<\st n|\frac{\d}{\d \tau}|\st n\> + \<\st n|H|\st n\>\right] d\tau
\]
where $S_E$ is the Euclidean action for spin coherent states which reproduces the Schr\"odinger equation with $\tau = -i t$, and $\mathcal{P}$ is the set of paths satisfying the boundary conditions $\mathbf{n}(0) = \mathbf{n}_1$ and $\mathbf{n}(T) = \mathbf{n}_2$.  

The dominant contribution to the path integral comes from instanton trajectories $\{\mathbf{n}(\tau) = (\theta(\tau),\phi(\tau))\}$, which are found as solutions to the Euler-Lagrange equations, and they depend on $s$, the cost function $H_p$, and the boundary conditions $\mathbf{n}_1, \mathbf{n}_2$.  The action integral can be reparametrized over $\theta$ instead of $\tau$, while the solution to the Euler-Lagrange equations can be used to eliminate $\phi$ and compute the total Euclidean action $S_I$ of the instantons,
\begin{equation}
 S_{I} = \frac{n}{2}\int_{\theta_1}^{\theta_2} \arccosh\left[1 + \frac{U(\theta, s^*) - U(\theta_1, s^*)}{J(1-s^*)\sin\theta}\right] \sin \theta \d \theta.\label{eq:essnaught}
\end{equation}
Although the instantons form the dominant contribution to the path integral, there will be corrections from the neighborhood of the instanton paths (sometimes called a tunneling pre-factor), and from multi-instanton paths~\cite{Co88}.  In Appendix \ref{app:instanton} the two methods for computing $\mathcal{U}_{21}$ are related to find,
\begin{equation}
\gap = \poly(n) e^{-S_I(n)}. \label{eq:instantonGapResult}
\end{equation}
Note that $S_I = S_I(n)$ may in general scale superlogarithmically with $n$ (due to the dependence of $U(\theta,s^*)$ on $n$), which allows the method to predict a variety of superpolynomial scalings for the gap.   In the next section we find a polynomial scaling of $S_I(n)$ with $n$ for spike Hamiltonians.  
\subsubsection{Application to spike Hamiltonians}\label{subsubsec:spikeinstantons}
\begin{figure}[ht]
 \centering
 \begin{subfigure}[t]{0.35\textwidth}
  \includegraphics[width=\textwidth]{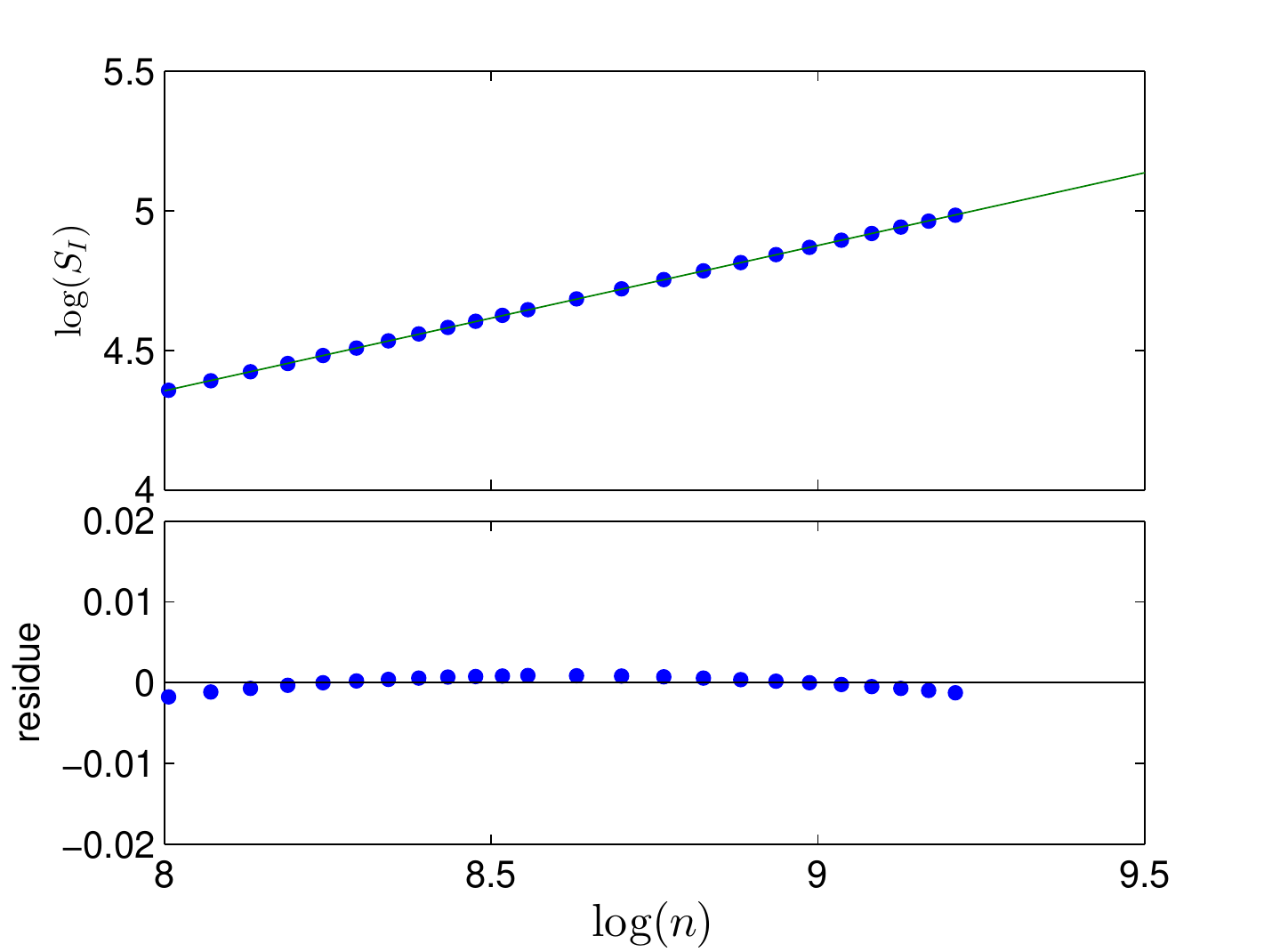}
  \caption{}
  \label{fig:powerlaw}
 \end{subfigure}
 \begin{subfigure}[t]{0.35\textwidth}
  \includegraphics[width=\textwidth]{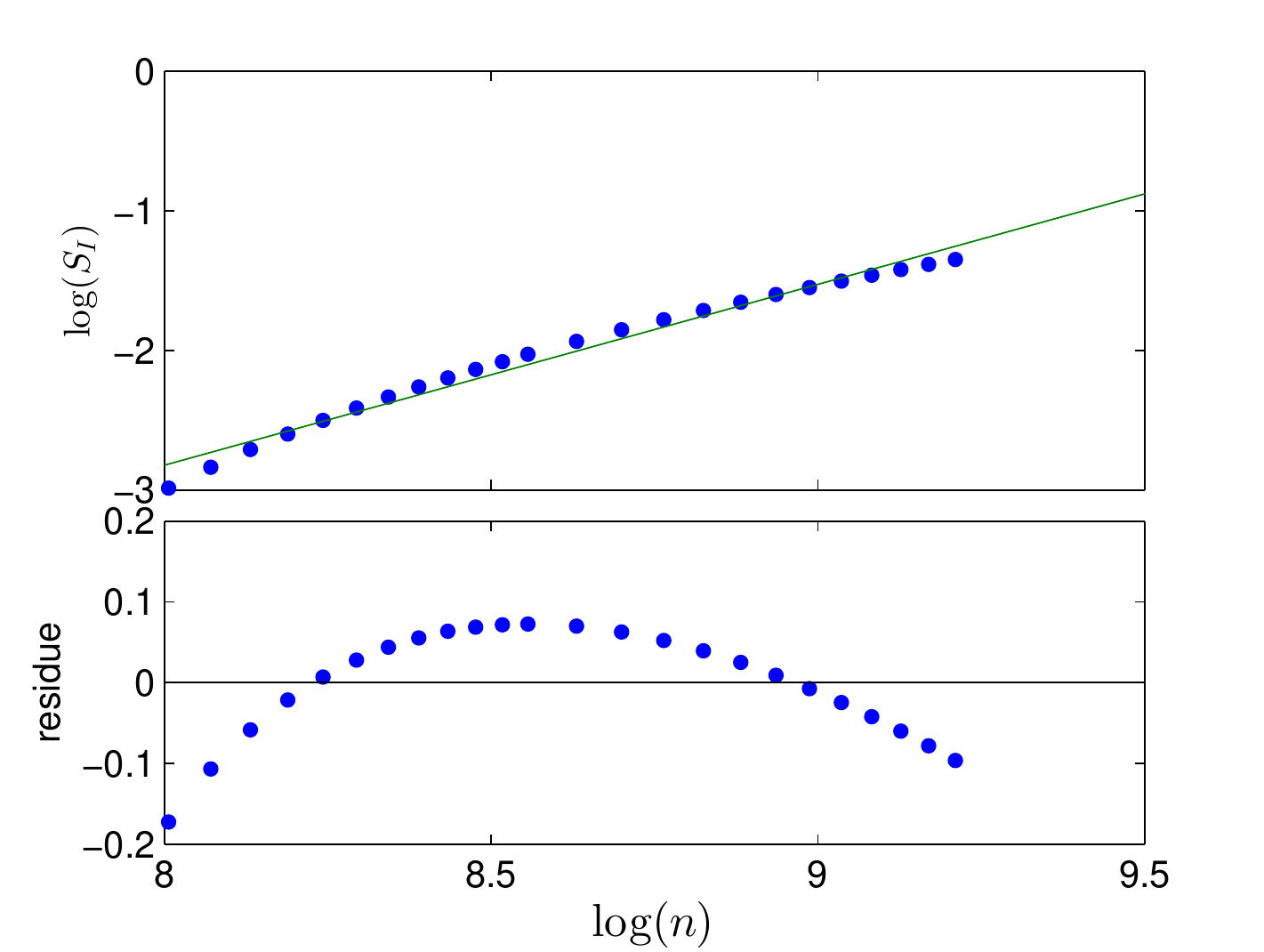}
  \caption{}
  \label{fig:powerlaw2}
 \end{subfigure}
 \caption{The curves of $\log(S_I)$ vs $\log(n)$ for different parameters. Fig.~\ref{fig:powerlaw} is given by parameters $\alpha = 1, \beta = 0.5$, and the residue is much smaller, but still concave. Fig.~\ref{fig:powerlaw2} is given by parameters $\alpha = 0.5, \beta = 0.2$, which is obviously concave and has large residue.  Note that the residue in Fig.~\ref{fig:powerlaw2} is over an order of magnitude larger than that of Fig.~\ref{fig:powerlaw}.}
\end{figure}

Numerically evaluating $S_I$ for the spike Hamiltonian, we find a power law scaling of $S_I$ with $n$ for most values of $(\alpha,\beta)$ where the method is applicable.  An example from the power law regime is shown in Fig.~\ref{fig:powerlaw} with $\alpha = 1, \beta = 0.5$, where the linear fit of $\log(S_I)$ vs $\log(n)$ has a small (but slightly concave) residue.  In contrast, for some values such as $\alpha = 0.5, \beta = 0.2$ in Fig.~\ref{fig:powerlaw2}, attempting to fit the data to a linear trend implies a large concave residue, and we conclude that $S_I$ grows more slowly than $n^c$ for any $c > 0$.
\begin{figure}[ht]
 \centering
 \includegraphics[trim = 0 0 0 40,width=0.7\textwidth]{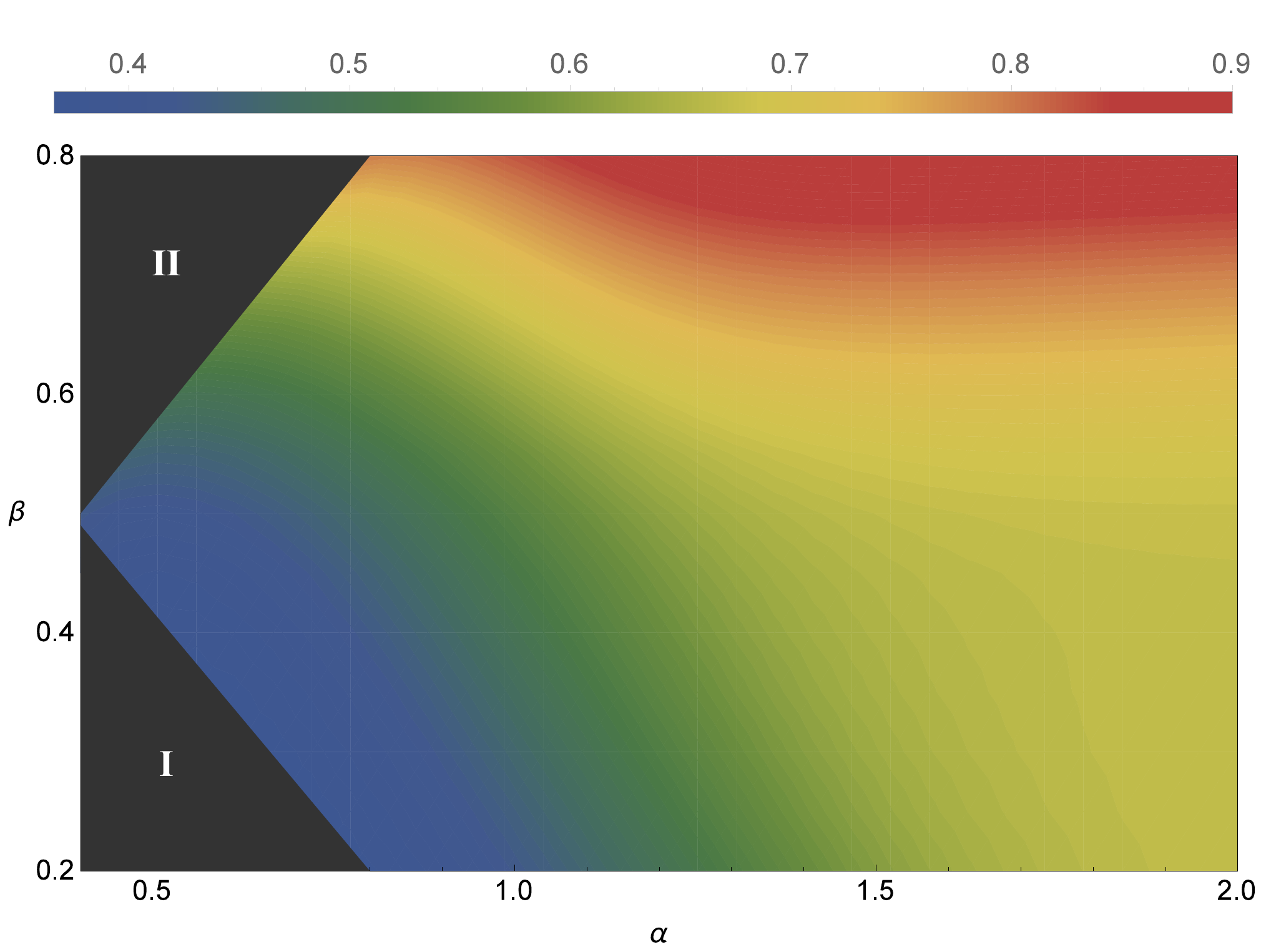}
 \caption{the slope of $\log(S_I)$ vs $\log(n)$ in a region of the $(\alpha, \beta)$ plane.}
 \label{fig:contour}
\end{figure}
A power law scaling of $S_I(n)$ implies a superpolynomially decreasing spectral gap, of the form $\gap(n) = \mathcal{O}(\poly(n)e^{-b n^c})$ for some constants $b,c > 0$.  The value of $c$ is determined by the instanton method, shown in Fig.~\ref{fig:contour} for a region of $(\alpha,\beta)$ plane, but the value of $b$ is not predicted by the method.  This complicates the matter of quantitative comparison between the instanton method and numerical diagonalization of the spike Hamiltonian, since an accurate determination of $b$ and $c$ from numerical diagonalization would require large system sizes and extremely precise determination of the gap size.   This contrasts with the relative ease of comparing the instanton method with numerical diagonalization for a cubic cost function~\cite{FGG02}, since in that case $S_I(n) = \Theta(n)$ so the scaling of the gap can be numerically fit to an ordinary exponential.

The power law scaling of $S_I$ with system size is shown in Fig. \ref{fig:contour}. For small $\alpha$ the exponent is positively correlated 
with both $\alpha$ and $\beta$. However, when $\alpha$ becomes large, the exponent remains almost the same for increasing $\alpha$.

Data is omitted in Regions I and II in Fig.~\ref{fig:contour} because the predictions of the instanton method are difficult to extract in these regions.  In Region I, $\alpha$ and $\beta$ are both small and the relationship between $\log(S_I)$ and $\log(n)$ is not linear.  Similarly, in Region II when $\beta$ is large while $\alpha$ is small, the spin coherent potential barrier between the local minima distorts and forms a pit which is even lower than the original, so the formula \eqref{eq:essnaught} no longer applies.

\subsection{The discrete WKB method} \label{subsec:WKB}
When the WKB method is applied to quantum systems with a discrete basis, the spatial derivatives are replaced finite differences, forming a recursive set of equations that determine an approximate wave function.  
As usual for the WKB method, it is natural to work with a Hamiltonian $H_{\nneg}(s)$ which is the opposite of $H(s)$,
\[ H_{\nneg}(s) \equiv \sin\theta X + \cos\theta (Z - H_{\spp}),
\]
and we will seek to estimate the eigenvalue gap between the two highest eigenstates of $H_{\nneg}$.  As before we treat the case $s = s^*$ and write $H_{\nneg} \equiv H_{\nneg}(s^*)$.

Let the highest eigenstate be $|\psi\>$ and define its wavefunction to be $C_k = \<k|\psi\>$. Since $H_{\nneg}$ is tridiagonal, $C_k$ satisfies a three-term recursion,
\begin{equation}
 p_j C_{j-1} + (w_j - E) C_j + p_{j+1} C_j = 0, \label{eq:rec}
\end{equation}
where we have introduced the new notations,
\[
p_j \equiv \frac{\sin\theta}{2}\sqrt{j(n+1-j)} \; \; , \; \; w_j \equiv \cos\theta(n/2 - j) +
 \begin{cases}
  3n^\alpha/4 & n/4 - n^\beta / 2 < j < n/4 + n^\beta / 2 \\
  0 & o.w,
 \end{cases}.
\]
and $E$ is any eigenvalue of $H_{\nneg}$. Note that the definition for $p_j$ and $w_j$ is still valid when $j$ is not integer. We assume $0 < \alpha < 1$ and $0 < \beta < 1/2$ for the rest of our calculation.  To simplify the recurrence in Eq.~\eqref{eq:rec}, we will omit the index $j$ and write $p_{j+1/2}$ as $p$ and $w_j$ as $w$ when there is no room for confusion.  Define the additional notations,
\[ U^+ \equiv w + 2p, \quad U^- \equiv w - 2p, \quad B \equiv \frac{E - w}{2p}, \quad v \equiv \sqrt{(U^+ - E)(E - U^-)}.
\]
The classically allowed region is the set of $j$ satisfying the inequality,
\[ U_j^- \leq E \leq U_j^+ ,
\]
and points $j_0$ for which $U_{j_0}^+ = E$ are called turning points.  If $w$ and $p$ are smooth functions of $j$, then the usual version of discrete WKB~\cite{Br93} approximates the wave function in the region of a turning point by
\begin{equation} C_j = 
 \begin{cases}
  \frac{1}{\sqrt v}\cos\left(\int_{j_0}^j \arccos B(j') \d j' - \pi / 4\right) & j > j_0 \\
  \frac{1}{2\sqrt{|v|}} \exp\left(-\int_j^{j_0} \operatorname{arccosh} B(j') \d j'\right) & j < j_0
 \end{cases}, \label{eq:formula}
\end{equation}
(note that \eqref{eq:formula} assumes $\d U^+ / \d j > 0$ at $j_0$, with a similar result for the other case).  
A challenge in our case is that $w$ is not smooth, due to the spike. In \cite{Am14} the continuous WKB method is adapted to an abruptly varying potential by replacing the WKB connection formula with the ``traditional'' boundary condition, i.e. the wave function and its derivative should be continuous at the turning point.  In the following subsections we will apply the traditional boundary condition to the spike Hamiltonian, but first we characterize the spectral gap in terms of properties of the approximate wave functions.

\subsubsection{Estimating the gap from approximate wave functions}

\begin{figure}[ht]
\centering
\includegraphics[width=0.5\textwidth]{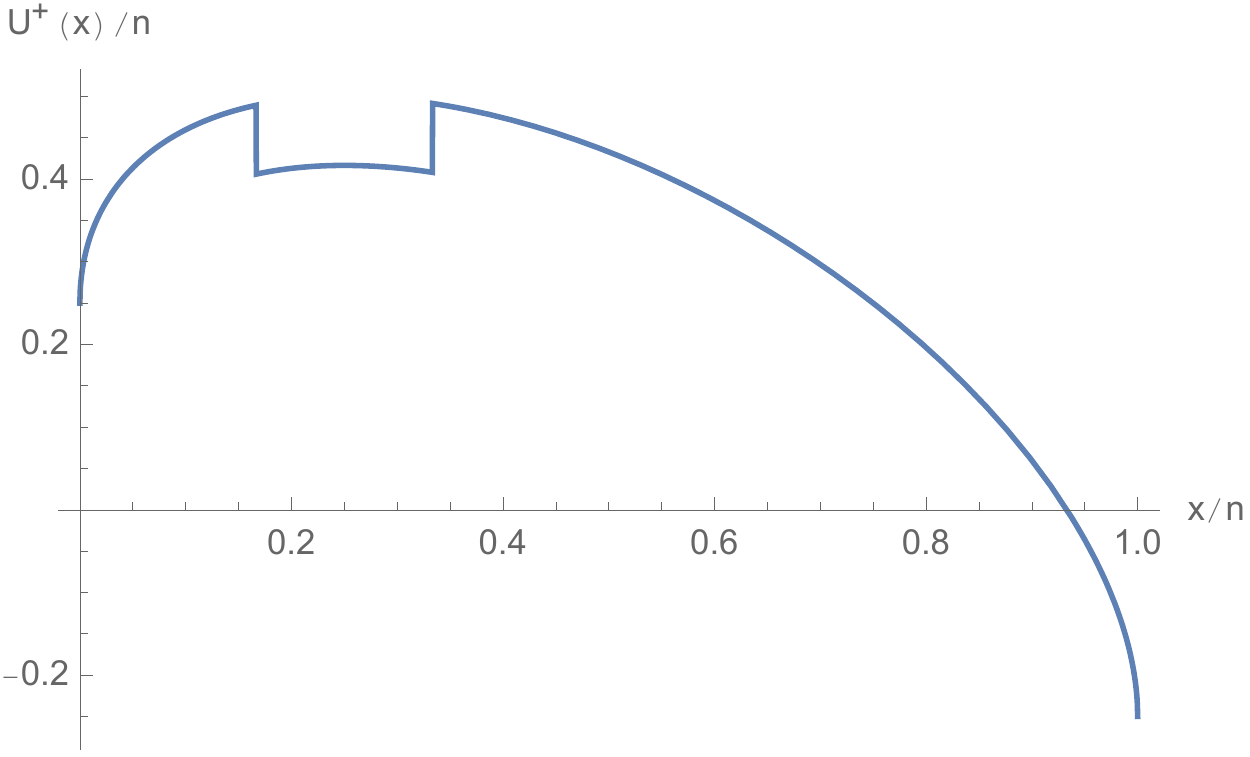}
\caption{The potential $U^+$ for $H_{\nneg}$.}
\label{fig:potential}
\end{figure}

The spectral gap in our case can be estimated using an argument similar to that found in~\cite{Ga98}.  The shape of the function $U^+$ is shown in Fig.~\ref{fig:potential}.

For $E$ high enough, there will be two localized (approximate) eigenstates located at the two largest 
maxima of $U^+$.  Just as in the previous section on the instanton method, the symmetric and antisymmetric superpositions of these states will have slightly different energy, which is the cause of the small gap. Let $\{C_j\}$ and $\{D_j\}$
be the two localized wave functions on the left and right, and define symmetric and antisymmetric wave functions $\{s_j\}$ and $\{a_j\}$ by $s_j \equiv (C_j + D_j) / \sqrt 2, a_j \equiv (C_j - D_j) / \sqrt 2.$ Let $E_1$ and  $E_0$ be the eigenvalues of $\{s_j\}$ and $\{a_j\}$, and from Eq.~\eqref{eq:rec} we have
\begin{align}
 (w_j - E_1) s_j + p_{j-1} s_{j-1} + p_j s_{j+1} &= 0, \label{eq:rec1}  \\
 (w_j - E_0) a_j + p_{j-1} a_{j-1} + p_j a_{j+1} &= 0. \label{eq:rec2}
\end{align}
Multiplying Eq.~\eqref{eq:rec1} by $a_j$ and subtracting Eq.~\eqref{eq:rec2} times $s_j$, and summing over $j \ge n/4$ yields
\begin{equation} (E_0 - E_1)\sum_{j \ge n / 4} s_j a_j =   p_{n/4-1}(a_{n/4}s_{n/4-1} - s_{n/4}a_{n/4-1})  .
\end{equation}
Since $\{C_j\}$ and $\{D_j\}$ are normalized and have approximately the same magnitude (by locality), the sum over $j$ is $\mathcal{O}(1)$, so this becomes
\begin{equation}
 \operatorname{gap}(n) = \Theta(n) (D_{n/4}C_{n/4-1} - C_{n/4}D_{n/4-1}). \label{eq:gap}
\end{equation}

\subsubsection{Appying the boundary conditions}
Since $\alpha < 1$ and $\beta < 1/2$, we can assume $E = (n + 1) / 2 - d$ with d = $\Theta(1)$, where $(n + 1) / 2$ is the maximum of $U^+$ in the spikeless Hamiltonian. Later we will calculate $d$ and see that is $\Theta(1)$.  

As discussed previously, we apply the WKB connection formula~\eqref{eq:formula} at the smooth turning point, together with the condition resulting from traditional boundary conditions at the abrupt turning point.
The smooth turning point is given by $U^+ = E$, which is
\[j_1 = \frac{n}{4}+\frac{d}{2}-\frac{1}{4}-\frac{1}{2}\sqrt{3d(n+1-d)},
\]
and the abrupt turning point is
\[j_2 = \frac{n}{4} - \frac{n^\beta}{2}.
\]

\newcommand{\Ca}{C^{(1)}}
\newcommand{\Cb}{C^{(2)}}
\newcommand{\Cc}{C^{(3)}}

Let
\[C_j =
 \begin{cases}
  \Ca_j & j \le j_1 \\
  \Cb_j & j_1 < j < j_2 \\
  \Cc_j & j \ge j_2
 \end{cases}.
\]
Then we know that
\begin{align}
 \Ca_j &= \frac{A}{2\sqrt{|v|}}\exp\left(-\int_j^{j_1}\arccosh B \d j'\right), \\
 \Cb_j &= \frac{A}{\sqrt v}\cos\left(\int_{j_1}^j\arccos B \d j' - \frac{\pi}{4}\right), \\
 \Cc_j &= \frac{B}{\sqrt{|v|}}\exp\left(-\int_{j_2}^j \arccosh B \d j'\right).
\end{align}
Note that a factor of 1/2 is absorbed in the definition of $B$ for $\Cc_j$. We do not consider the abrupt change at $n/4 + n^\beta/2$, as we assume
the amplitude is already small enough at that point. The validity of this assumption will be discussed later.

The connection condition at $j_2$ requires
\[ \left.\frac{1}{\Cb_j}\frac{\d \Cb_j}{\d j}\right|_{j = j_2^-}
 = \left.\frac{1}{\Cc_j}\frac{\d \Cc_j}{\d j}\right|_{j = j_2^+},
\]
which simplifies to
\begin{equation} \left.-\frac{1}{2} \frac{v'}{v}\right|_{j = j_2^-} - \tan\left(\int_{j_1}^{j_2}\arccos B \d j' - \frac{\pi}{4}\right) \left. \arccos B \right|_{j = j_2^-}
= \left.-\frac{1}{2} \frac{|v|'}{|v|}\right|_{j = j_2^+} - \left.\arccosh B\right|_{j = j_2^+}.
\label{eq:connection}
\end{equation}
For large $n$ the leading behavior is,
\begin{align*}
 \left.\frac{1}{2} \frac{|v|'}{|v|}\right|_{j = j_2^+} - \left.\frac{1}{2} \frac{v'}{v}\right|_{j = j_2^-} &= \Theta(n^{b-1}), \\
 \left.\arccosh B\right|_{j_2^+} &= \Theta(n^{(\alpha - 1) / 2}), \\
 \left.\arccos B\right|_{j_2^-} &= \Theta(n^{-1/2}).
\end{align*}
If we only keep the terms with highest order in $n$, we can approximate the integral
\[
 \int_{j_1}^{j_2}\arccos B \d j' \approx \frac{d\pi}{2},
\]
so~\eqref{eq:connection} yields
\begin{equation}
 \tan(d\pi/2 - \pi / 4) = \Theta(n^{\alpha / 2}). \label{eq:tangent}
\end{equation}
Since $n^{\alpha / 2} \to +\infty$ for $n \to \infty$, the solution would be $d\pi/2 - \pi / 4 = \pi / 2 - o(1)$, and $d = 3/2 - o(1)$ (the second $o(1)$
might be negative, as we ignored some terms when calculating the integral). This means the
largest eigenvalue of $H_{\nneg}$ is approximately the second largest eigenvalue for the spikeless Hamiltonian.

\subsubsection{Estimating the gap}\label{subsubsec:spikeGapWKB}
Following \cite{Fu47}, we normalize the WKB wave function using only the part in classically allowed region.  This is intuitively reasonable, since the wave function decays faster than any polynomial in the classically forbidden region. We approximate $\Cb$ to leading order in $n$, and the integral is $\Theta(1)$, which indicates
that $A = \Theta(1)$. The connection formula at $j_2$ requires,
\[
 \frac{B}{A} = \frac{\left.\sqrt{|v|}\right|_{j = j_2^+}}{\left.\sqrt{v}\right|_{j = j_2^-}}\cos\left(\int_{j_1}^{j_2}\arccos B \d j' - \pi / 4\right) 
 = \Theta(n^{-\alpha/4}),
\]
where the last step is evaluated using~\eqref{eq:tangent}.

It can be verified that $\{C_j\}$ and $\{D_j\}$ are symmetric w.r.t. $j = n/4$, up to the highest order in $n$.  Therefore we can evaluate the difference in~\eqref{eq:gap} as a derivative,
\[
 \operatorname{gap}(n)= -\Theta(n) C_{n/4}\left.\frac{\d C_j}{\d j}\right|_{j = n / 4} 
 = \Theta(n^{-\alpha/2 + 1})\frac{1}{|v|}\exp\left(-2\int_{j_2}^{n/4}\arccosh B \d j'\right)
 \left(\frac{|v|'}{2|v|} + \arccosh B\right),
\]
which then simplifies to
\[
 \operatorname{gap}(n)
 = \Theta(n^{-\alpha/2})\exp\left(-2\int_{j_2}^{n/4}\arccosh B \d j'\right).
\]
Finally, the integral in the exponent evaluates to
\[
 \int_{j_2}^{n/4}\arccosh B \d j' = \Theta(n^{\alpha/2 + \beta - 1/2}).
\]
Therefore when $\alpha + 2\beta > 1$, the discrete WKB method suggests that gap will be superpolynomially small, which matches the numerical result in \ref{subsec:numerics} and is also generally consistent with the predictions from the spin coherent instanton method in Fig.~\ref{fig:contour}.  However, the method breaks down at small $\beta$ by giving a result which is independent of $\beta$, and this result is to be expected since a system with a narrow spike is unlikely to satisfy the initial assumption that $\{C_j\}$ and $\{D_j\}$ are localized.  

\section{Diabaticity in the spike of width 1}
\label{sec:diabatic}
When the adiabatic algorithm is run diabatically, the quantum state no longer follows the instantaneous ground state of the evolving Hamiltonian. However, recently there has been an increased interest in advantageous diabatic effects~\cite{MAL15-2,MAL15,CFL+14,TIM+12}, which for some cost functions can produce a modest success probability in much less time than would be predicted by the adiabatic theorem.  In particular, the spike problem with width 1 has been shown to benefit from a diabatic cascade phenomenon \cite{MAL15}.

In this section we will apply the intuition gained from the ground state ansatz in Section \ref{sec:width1} to understand the location of the avoided crossings in the spike system.  We compute a recursion relation involving the spikeless eigenstates, and use it to show that the value of the adiabatic parameter at which the first avoided crossing occurs between levels $k$ and $k-1$ is a monotonically decreasing of $k$.  In other words, the sequence of initial avoided level crossings starts at the top of the energy spectrum and proceeds down to the lowest energy levels in consecutive order.   This sequence is only a part of the larger pattern in the avoided crossings which could explain the diabatic cascade, but it may be useful for understanding the full pattern in future work.

In Section \ref{sec:width1} we saw that the value of the adiabatic parameter at the critical point, $s = s^* \equiv (\sqrt 3-1)/2$, is the same value of $s$ at which the first excited state probability distribution of the spikeless system has a node on the state $|n/4\rangle$ (the node of a sequence is the index at which the sequence becomes 0 or changes sign).  The key observation of this section is that a similar connection between nodes in the spikeless system, and avoided crossings in the spike system, also applies between the higher energy levels. 

\begin{figure}[ht]
 \centering
  \includegraphics[width=0.7\textwidth]{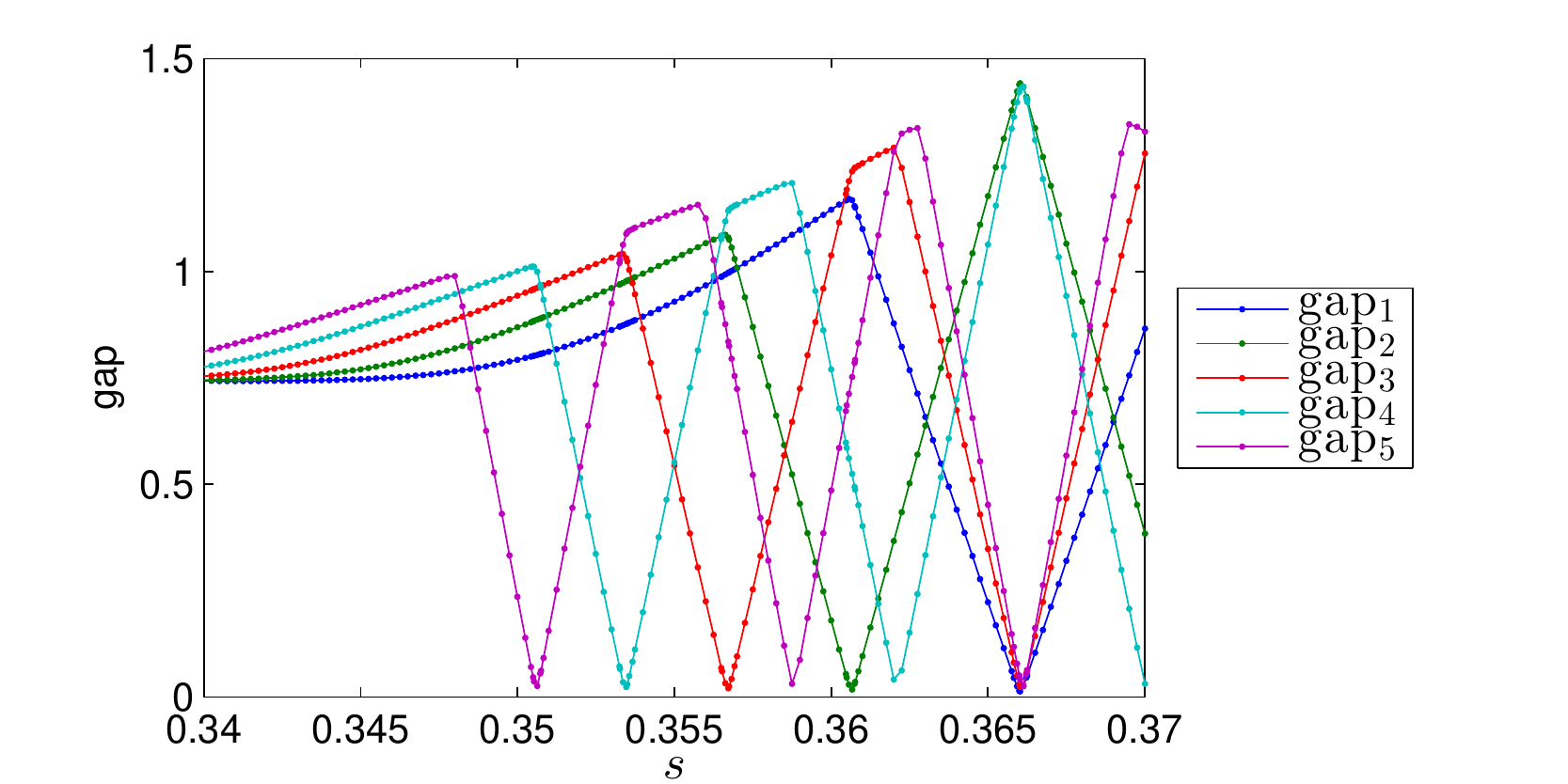}
 \caption{The evolution of the $\gap_t \equiv E'_t - E'_{t-1}$, $t = 1, 2, \ldots, 5$ when $n=10000$. 
}
 \label{fig:gap}
\end{figure}

Label the energy eigenstates of the spikeless system with size $n$ in order of increasing energy $\{|\psi_t^{(n)}\> : t= 0,\ldots ,n\}$. We write the eigenvalues of the spike system as $\{E_t'(s): t = 0, \ldots, n\}$, and define
\[
 \gap_t(s) = E_t'(s) - E_{t-1}'(s).
\]
Let $s^i_t$ be the value of $s$ at which the $i$-th node of $|\psi_t^{(n)}\>$ is at $n/4$ (by nodes of $|\psi_t^{(n)}\>$ we mean nodes of the sequence $\<k|\psi_t^{(n)}\>$). In general, $s^i_t$ is a function of $n$, but we will leave this dependence implicit in our notation.  We observe that the spike system has an avoided crossing between levels $t$ and $t-1$, 
\begin{equation}
\lim_{n\to\infty} \gap_t(s^i_{t}) = 0 
\end{equation}
for each $i = 1,\ldots, t-1$.  Numerical evidence for the conjecture is presented in Fig. \ref{fig:gap}, which shows that for $n = 10000$, $\operatorname{gap}_t$ becomes tiny at the predicted values $s^i_t$. Numerical calculation further shows that the gap is decreasing as an inverse polynomial in the system size.  

Now that we have established a correspondence between the location of these avoided crossings in the spike system and the nodes in spikeless eigenstates, the rest of this section will demonstrate that the sequence of first nodes $\{s_t^1\}$, denoted $\{s_t\}$ for short, is monotonically decreasing with $t$, i.e. $s_{t+1} < s_t$.  This implies that the first avoided crossing happens between the top two energy levels, and so on cascading down through the rest of the eigenstates in consecutive order.

\subsection{Ordering of the nodes}
First we compute an expression for the spikeless wavefunctions in the symmetric subspace, and then we establish a recurrence relation that demonstrates the claimed relations between the nodes.  In the full Hilbert space of dimension $2^n$, the ground state for each qubit is $ \cos(\theta/2)|0 \>+\sin(\theta/2)|1 \>$ and the excited state is
$ \sin(\theta/2)|0 \>-\cos(\theta/2)|1 \>$. Therefore $|\psi_t^{(n)}\>$ in the symmetric subspace is just the uniform superposition of all tensor product states
with $t$ excited qubits and $n-t$ ground-state qubits.  From this we obtain,
\begin{equation}
 \<k|\psi_t^{(n)}\> = \sqrt\frac{\binom{n}{t}}{\binom{n}{k}} \tan^{t+k}(\theta/2)\cos^n(\theta/2)\sum_{j=0}^t
 \left(-\cot^2(\theta/2)\right)^j\binom{t}{j}\binom{n-t}{k-j}.  \label{eq:eigenstate}
\end{equation}
Note that this expression is valid even for $k < 0$ or $k > n$, where the result is simply 0.
With some manipulation of binomial coefficients, we can establish the recurrence relation
\[
 P_{n+1,k}\<k|\psi_{t+1}^{(n+1)}\> \propto 
 P_{n,k}\<k|\psi_t^{(n)}\>-
 P_{n,k-1}\<k-1|\psi_t^{(n)}\>,
\]
where
\[
 P_{n,k} \equiv \sqrt{\binom{n}{k}}\sin^k(\theta/2)\cos^{n-k}(\theta/2),
\]
and the constant of proportionality is independent of $k$. This means that the $(t+1)$-th excited state wave function can be obtained through the difference of
a sequence related to the $t$-th excited state wave function.


We want to prove that for the same $s$, the first node of $(t + 1)$-th energy level is smaller than the first node
of $t$-th. This will lead to the desired conclusion $s_{k+1} < s_k$.

Define two sequences,
\[
S = \left\{P_{n+1,k}\<k|\psi_{t+1}^{(n+1)}\>\right\}_{k=0}^{n+1} \quad , \quad S' = \left\{P_{n,k}\<k|\psi_{t}^{(n)}\>\right\}_{k=0}^n.
\]
By the recurrence relation derived above we see that $S$ is given by the difference of consecutive points of $S'$.   From Eq.~\eqref{eq:eigenstate} we can see that the first element of the sequence $S'$,
which is proportional to $\<0|\psi_{t}^{(n)}\>$, is positive. We consider the smallest node $x_0$ for $S'$, if any.
Then the sequence element of $S'$ at $x_0$ is non-positive,
so the difference of the sequence $S'$ should become non-positive somewhere before $x_0$. This proves that the first node of 
$\<k|\psi_{t+1}^{(n+1)}\>$ is smaller than the first node of $\<k|\psi_{t}^{(n)}\>$.
When we go from $n+1$ to $n$, the node cannot become larger. So
the first node of $\<k|\psi_{t+1}^{(n)}\>$ is also smaller than the first node of $\<k|\psi_{t}^{(n)}\>$.

Considering the indices of the nodes of the wave function increase as $s$ increases, we have $s_{t+1} < s_t$.
\section{Outlook}
One of the most significant challenges in finding applications for quantum annealing is to understand how the spectral gap that determines the adiabatic run-time depends on the details of the cost function.  Bit-symmetric QA Hamiltonians are an increasingly important test bed for new insights into quantum annealing, which is why we have developed multiple methods for approximating and in some cases rigorously bounding the spectral gap for a class of spike Hamiltonians.  

The development of a sufficiently accurate ansatz for the ground state of the width 1 spike was the key to generalizing Reichardt's rigorous variational-comparison bound to the case when the system has a critical point.   The anzatz also led us to understand that avoided crossings in the spike system correspond to nodes of the spikeless eigenfunctions crossing the spike.  This may provide useful intuition for the location and scaling of the spectral gap when a Hamiltonian has a barrier inserted at the position of an eigenfunction node.  

In addition to supplying calculational details for spin coherent instantons which may be of use in future studies, we clarify the nature of the method: it is essentially an ansatz for approximate wave functions, because it assumes a form of ground state and first excited state as symmetric and antisymmetric combinations of localized states.  The instanton method has the advantage that it is fairly easy to compute the Euclidean action for potentials of various shapes, but the disadvantage is that the coarseness of the approximation means it only yields a rough estimate of the spectral gap.  Better estimates were obtained using the discrete WKB method, which more explicitly constructs approximate wave functions.

Finally, the method we give for identifying the locations of the avoided crossings between the higher energy eigenstates of the spike of width 1 could lead to a tractable analysis of the pattern of crossings, which we hope will be of use in furthering the understanding of the diabatic cascade phenomenon.  

\section*{Acknowledgement}
We are very grateful to Aram Harrow for guiding us at many points throughout this work.  We thank Jeffrey Goldstone for discussing the application of spin coherent states in~\cite{FGG02} and for encouraging us to recover the calculation in Appendix \ref{app:instanton}.  We also thank David Gosset for suggesting the use of Lemma \ref{lem:lb} to lower bound the ground state energy, and we thank Bill Kaminsky for referring us to exposition of coherent spin path integrals in \cite{Ga00}.  Linghang Kong acknowledges the Tsinghua-MIT CTCS undergraduate exchange program, through which he visited MIT in Spring 2015 and did the major part of this work. He also thanks the Undergraduate Research Opportunity Program in MIT.   Elizabeth Crosson gratefully acknowledges funding provided by the Institute for Quantum Information and Matter, an NSF Physics Frontiers Center (NSF Grant PHY-1125565) with support of the Gordon and Betty Moore Foundation (GBMF-12500028), and is also grateful for support received while completing a portion of this work at the MIT Center for Theoretical Physics with funding from NSF grant number CCF-1111382.

\appendix
\section{Variational calculation}
\label{app:variational}
The ground state of the spikeless Hamiltonian $H$ is
\begin{equation}
 |\psi_0\> = \sum_{k=0}^n \sqrt{\binom{n}{k}} \left(\frac{1}{2}\right)^{k}\left(\frac{\sqrt 3}{2}\right)^{n - k}\left|k\right\>, \label{eq:ampl-g}
\end{equation}
and the first excited state wave function for the spikeless Hamiltonian is
\begin{align}
 |\psi_1\> 
 =& \sum_{k=0}^n \sqrt\frac{\binom{n}{k}}{3n}
 \left(\frac{1}{2}\right)^{k}\left(\frac{\sqrt 3}{2}\right)^{n - k}\left(n - 4k\right)\; \left|k\right>. \label{eq:ampl-e}
\end{align}

The spikeless Hamiltonian $H$ is the negative of the angular momentum operator $\frac{\sqrt 3}{2}X + \frac{1}{2}Z$, which implies
\begin{equation*}
 \<\psi_0|\frac{\sqrt 3}{2}X + \frac{1}{2}Z|\psi_0\> = \frac{n}{2}\quad,\quad\<\psi_1|\frac{\sqrt 3}{2}X + \frac{1}{2}Z|\psi_1\> = \frac{n}{2} - 1.
\end{equation*}
Now we consider the expected value of $X$ and $Z$ of the state $|\psi_{\abs}\>$ defined in Eq.~\eqref{eq:abs1prime}.
The expected value of the $Z$ is unaffected by taking the absolute value of the amplitudes since it is diagonal,
\begin{equation}
 \<\psi_{\abs}|Z|\psi_{\abs}\> = \sum_{k=0}^n\left(J - k\right) (\<k|\psi_{\abs}\>)^2 = \sum_{k=0}^n\left(J - k\right) (\<k|\psi_1\>)^2 = \<\psi_1|Z|\psi_1\>\nonumber
 \end{equation}
 Define $A_{n,k} = \sqrt{k(n+k-1)}$.  The expected value of $X$ is also unaffected by taking the absolute value of the amplitudes since $\<n/4|\psi_1\> = \<n/4|\psi_{\abs}\> =0$,
 \begin{align}
 \<\psi_{\abs}|X|\psi_{\abs}\> =& \sum_{k=1}^n A_{n,k}\<k|\psi_{\abs}\>\<k-1|\psi_{\abs}\> \nonumber\\
 =& \sum_{k=1}^{n/4 - 1} A_{n,k} \<k|\psi_1\>\<k-1|\psi_1\> - A_{n,n/4}\<n/4|\psi_1\>\<n/4-1|\psi_1\> \nonumber \\
 &- A_{n,n/4 + 1}\<n/4+1|\psi_1\>\<n/4|\psi_1\>
 + \sum_{k=n/4+2}^{n} A_{n,k}(-\<k|\psi_1\>)(-\<k-1|\psi_1\>) \nonumber \\
 =& \sum_{k=1}^n A_{n,k}\<k|\psi_1\>\<k-1|\psi_1\>  = \<\psi_1|X|\psi_1\>. \label{eq:sgn}
\end{align}
where step \eqref{eq:sgn} comes from $\<n/4|\psi_1\> = 0$ so that the terms for $k = n/4$ and $k = n/4+1$ can be negated.
We now carry out the variational method by finding the state $|\psi\> \in \spc S$ that minimizes ${\<\psi| H|\psi\>}/{\<\psi|\psi\>}$,
which also minimizes
\begin{equation*}
\frac{\<\psi| H|\psi\>}{\<\psi|\psi\>} - \<\psi_1| H_0|\psi_1\> \nonumber
 = \frac{\<\psi_{\abs}| H|\psi_{\abs}\> + 2x \<\psi_{\abs}| H|\psi_0\> + x^2\<\psi_0| H|\psi_0\>}{1 + 2x \<\psi_{\abs}|\psi_0\> + x^2} - \<\psi_{\abs}| H_0|\psi_{\abs}\>. \nonumber
\end{equation*}
Reducing this further,
\begin{align}
 =& \frac{\<\psi_{\abs}| H_0|\psi_{\abs}\> + 2x \<\psi_{\abs}| H_0|\psi_0\> + x^2\<\psi_0| H_0|\psi_0\> + \cos\theta x^2 \<\psi_0|H_{sp}|\psi_0\>}{1 + 2x \<\psi_{\abs}|\psi_0\> + x^2}
 - \<\psi_{\abs}| H_0|\psi_{\abs}\>, \label{eq:sgn2}\\[10pt]
 =& \frac{2x (\<\psi_{\abs}| H_0|\psi_0\> - \<\psi_{\abs}|\psi_0\>\<\psi_{\abs}| H_0|\psi_{\abs}\>) + 
 x^2(\<\psi_0| H_0|\psi_0\>-\<\psi_{abs}| H_0|\psi_{\abs}\>) +\cos\theta x^2 \<\psi_0|H_{sp}|\psi_0\>}
 {1 + 2x \<\psi_{\abs}|\psi_0\> + x^2}, \nonumber\\[10pt]
 =& \frac{2x (-J\<\psi_{\abs}|\psi_0\> + (J-1)\<\psi_{\abs}|\psi_0\>) + 
  x^2 (\cos\theta \<\psi_0|H_{sp}|\psi_0\>-1)}{1 + 2x \<\psi_{\abs}|\psi_0\> + x^2}, \nonumber \\[10pt]
 =& \frac{-2x \<\psi_{\abs}|\psi_0\> + 
  x^2 (\cos\theta \<\psi_0|H_{sp}|\psi_0\>-1)}{1 + 2x \<\psi_{\abs}|\psi_0\> + x^2}, \label{eq:target}
\end{align}
where step \eqref{eq:sgn2} again uses the fact that $\<n/4|\psi_{\abs}\> = 0$. It remains to calculate $\<\psi_{\abs}|\psi_0\>$,
\[
 \<\psi_{\abs}|\psi_0\> = \sum_{k=0}^n \<k|\psi_0\>\<k|\psi_{\abs}\>
 = \frac{1}{\sqrt{3n}}\sum_{k=0}^n \binom{n}{k} \left(\frac{1}{4}\right)^{k}\left(\frac{ 3}{4}\right)^{n - k} |n - 4k|. 
\]
According to the formula for mean deviation of binomial distribution given in \cite{Fr45}, we have
\begin{align*}
 \sum_{k=0}^n \binom{n}{k}\left(\frac{1}{4}\right)^k \left(\frac{3}{4}\right)^{n-k} \left|k-\frac{1}{4}n\right| &= 2n \times \frac{1}{4} \times \frac{3}{4} \max_{0\le k\le n-1} \binom{n-1}{k}\left(\frac{1}{4}\right)^k \left(\frac{3}{4}\right)^{n-1-k} \\[7pt]
 =& \frac{3n}{8} \binom{n-1}{n/4 - 1}\left(\frac{1}{4}\right)^{n/4 - 1} \left(\frac{3}{4}\right)^{3n/4} \\[7pt]
 =& \frac{3n}{8} \binom{n}{n/4}\left(\frac{1}{4}\right)^{n/4} \left(\frac{3}{4}\right)^{3n/4},
\end{align*}
so
\[
 \<\psi_{\abs}|\psi_0\> = \frac{\sqrt{3n}}{2} \binom{n}{n/4}\left(\frac{1}{4}\right)^{n/4} \left(\frac{3}{4}\right)^{3n/4}=\sqrt{\frac{2}{\pi}}(1 + O(1/n)).
\]
Also, note that
\begin{equation*}
 \<\psi_0|H_{\spp}|\psi_0\> = \sqrt{\frac{8}{3\pi}} n^{\alpha - 1/2} (1 + O(1/n)).
\end{equation*}
If we take
\[
 x = x_0 \equiv \frac{\sqrt 3}{2} n^{1/2 - \alpha},
\]
and plug all these into Eq.~\eqref{eq:target}, we will have
\begin{equation*}
 \left.\left(\frac{\<\psi| H|\psi\>}{\<\psi|\psi\>} - \<\psi_1| H_0|\psi_1\>\right)\right|_{x=x_0} = - \sqrt{\frac{3}{2\pi}} n^{1/2 - \alpha} (1 + o(1)).
\end{equation*}
This means when $\alpha > 1/2$,
the average energy of trial wave function on $ H$ is $\Theta(n^{1/2 - \alpha})$ less than the energy of first excited state for $ H_0$, which provides the lower bound on the gap which is used in Theorem~\ref{thm:width1}.    

\section{Evaluating the Euclidean path integral for spin tunneling}
\label{app:instanton}
In this section we consider general bit-symmetric quantum annealing Hamiltonians,
\begin{equation}
 H(s) = (1-s)C_n \left(\frac{n}{2} - X\right) + s H_p, \label{eq:general-h}
\end{equation}
where $H_p$ is diagonal in the computational basis.  The spin coherent state $|\theta, \phi\>$ is related to the basis $\{|k\>\}^n_{k = 0}$ by 
\begin{equation*}
 |\theta, \phi\> = \sum_{k=0}^n \sqrt{\binom{n}{k}}\left(\cos\frac{\theta}{2}\right)^{n-k}\left(e^{i\phi}\sin\frac{\theta}{2}\right)^k|k\>.
\end{equation*}
The energy expectation $\<\theta,\phi|H_p|\theta,\phi\>$ is independent of $\phi$ because $H_P$ is diagonal, so we can express the energy of the spin coherent states in terms of the quantity $W(\theta, s) \equiv s\<\theta, \phi|H_p|\theta, \phi\>$,
\begin{align*}
E(\theta, \phi, s) \equiv \<\theta, \phi|H(s)|\theta,\phi\> = \frac{n}{2}C_n (1-s) (1-\sin\theta\cos\phi) + W(\theta,s),
\end{align*}

We also define
\[
 U(\theta, s) \equiv E(\theta, 0, s) = \frac{n}{2}C_n(1-s)(1 - \sin\theta) + W(\theta, s).
\]
Suppose $s = s^*$ is a particular value of the adiabatic parameter for which there are two local minima for $U(\theta, s)$. We label these minima $\theta_1$ and $\theta_2$  with $\theta_1 < \theta_2$, and as before we set $H \equiv H(s^*)$.  As explained in \ref{subsec:instanton} the goal is to evaluate the transition amplitude $\mathcal{U}_{2 1} = \<\theta_2, 0|e^{-T H}|\theta_1, 0\>$ by a path integral,
\begin{equation}
\mathcal{U}_{2 1} = \int_{\mathcal{P}} \mathcal{D}\st n(\tau) e^{-S_E[\st n(\tau)]} \quad , \quad  S_E[\st n(\tau)] = \int_{0}^{T} \d \tau \left[\<\st n|\frac{\d}{\d \tau}|\st n\> + \<\st n|H|\st n\>\right]. \label{eq:euc-act}
\end{equation}
In terms of the angles $|\st n\> = |\theta,\phi\>$, the Euclidean Lagrangian becomes
\begin{equation*}
\<\st n|\frac{\d}{\d \tau}|\st n\> + \<\st n|H|\st n\> = \<\theta,\phi |\left(\dot{\theta}\frac{\partial}{\partial \theta} + \dot{\phi}\frac{\partial}{\partial \phi} \right) | \theta,\phi\> + E(\theta,\phi,s),
\end{equation*}
therefore we compute
\begin{equation}
 \<\theta, \phi|\frac{\p}{\p \theta}|\theta,\phi\> = 0 \; \;, \; \; \<\theta, \phi|\frac{\p}{\p \phi}|\theta,\phi\> = i J(1 - \cos\theta),
\end{equation}
and Eq.~\eqref{eq:euc-act} in terms of the angles is
\[
S_E[\theta(\tau),\phi(\tau)] = \int_{0}^{T} d\tau \left[i J\dot{\phi}(1-\cos \theta) + E(\theta,\phi,s)\right].
\]
The main contribution to the tunneling amplitude $\mathcal{U}_{2 1}$ will come from the paths that minimize the Euclidean action.  Since the Euclidean action of these instantons is stationary under infinitesimal variations of the path, they can be found as solutions to the Euler-Lagrange equations,
\begin{align}
 iJ\dot\phi\sin\theta &= - \frac{\p E(\theta, \phi, s^*)}{\p \theta}, \label{eq:EL1} \\
 iJ\dot\theta\sin\theta &= \frac{\p E(\theta, \phi, s^*)}{\p \phi}. \label{eq:EL2}
\end{align}
The equations above show that one of $\dot\theta$ and $\dot\phi$ should be imaginary.
Since $\theta$ is the tunneling coordinate and should be real, we know that the real part of $\phi$ is invariant in the process. In the beginning $\phi = 0$, so $\phi$ is purely imaginary through the process.

Even in Euclidean time, the solutions to the Euler-Lagrange equations preserve energy (which can be seen by subtracting $\dot\theta$ times Eq.~\eqref{eq:EL1} by $\dot\phi$ times Eq.~\eqref{eq:EL2}), 
\begin{align*}
 \frac{\d}{\d \tau} E(\theta,\phi, s^*) =\left(\dot{\theta}\frac{\partial}{\partial \theta} + \dot{\phi}\frac{\partial}{\partial \phi} \right) E(\theta,\phi, s^*) = 0,
\end{align*}
which implies
\begin{equation}
 E(\theta(\tau), \phi, s^*) = E(\theta_1, 0, s^*). \label{eq:conserve}
\end{equation}
Eq.~\eqref{eq:conserve} gives the relation of $\phi$ and $\theta$,
\begin{equation*}
\cos\phi = 1 + \frac{U(\theta, s^*) - U(\theta_1, s^*)}{JC_n(1-s^*)\sin\theta}.
\end{equation*}
Considering $\phi$ is purely imaginary, we define
\begin{equation}
 \varphi(\theta) = \arccosh\left[1 + \frac{U(\theta, s^*) - U(\theta_1, s^*)}{JC_n(1-s^*)\sin\theta}\right], \label{eq:varphi}
\end{equation}
and we will have $\phi = \pm i \varphi(\theta)$ through the process. However, consider Eq.~\eqref{eq:EL2}, which further simplifies to
\begin{equation*}
 iJ\dot\theta = 2(1 - s^*) \sin\phi.
\end{equation*}
If $\phi = -i \varphi(\theta)$, $\dot\theta$ will always be negative, which contradicts $\theta_2 > \theta_1$. The only possible relation is
$\phi = i \varphi(\theta)$.

Then we consider the Euclidean action for these two paths. From Eq.~\eqref{eq:euc-act} we have
\begin{align*}
 S_E(\st n) &=
 \int_0^T \d \tau \left(iJ\dot\phi(1 - \cos\theta) + E(\theta, \phi, s^*)\right) \\
 &= T E(\theta_1, 0) + iJ \left(\phi(1-\cos\theta)\bigg|_0^T - \int_0^T \phi \frac{\d(1 - \cos\theta)}{\d \tau} \d \tau\right) \\
 &= T E(\theta_1, 0) - iJ  \int_{\theta_1}^{\theta_2} \phi \sin\theta \d\theta,\\
 &= T E(\theta_1, 0) + J  \int_{\theta_1}^{\theta_2} \varphi(\theta) \sin\theta \d\theta, \\
 &= T E(\theta_1, 0) + S_I,
\end{align*}
where $S_I$ is defined as
\begin{equation}
 S_I \equiv J\int_{\theta_1}^{\theta_2} \varphi(\theta) \sin\theta \d \theta = \frac{n}{2}\int_{\theta_1}^{\theta_2} \arccosh\left[1 + \frac{U(\theta, s^*) - U(\theta_1, s^*)}{JC_n(1-s^*)\sin\theta}\right] \sin \theta \; \d \theta \;, \label{eq:s0}
\end{equation}
as claimed in Section \ref{subsec:instanton}.
\subsection{Relating the tunneling amplitude to the gap}
Now we can describe the additional details~\cite{Ga98} that are necessary for relating $S_I$ to the tunneling amplitude $\mathcal{U}_{21}$, and for relating $\mathcal{U}_{21}$ to the spectral gap $\Delta E$ of the system.  As discussed in Section \ref{subsec:instanton}, the ground state $|\psi_+\>$ and first excited state $|\psi_-\>$ are expected to have the form
\[
|\psi_{\pm}\> = \frac{1}{\sqrt 2}\left(|\psi_1\> \pm |\psi_2\>\right) \quad , \quad \<\st n_i|\psi_j\> \approx a_i \delta_{i,j},
\]
where $a_i = \Omega(1)$ and the approximate equality means equality within $\poly(n^{-1})$ additive error.  In other words, $|\psi_j\>$ is localized and has large overlap with $|\mathbf{n}_j\>$, but is almost orthogonal with $|\mathbf{n}_{i}\>$ for $i \neq j$. Labeling the eigenvalues of $|\psi_{\pm}\>$ by $E_{\av} \pm \frac{1}{2}\Delta E$ and expanding $\mathcal{U}_{21}$ in terms of the energy eigenbasis,
\begin{equation}
\mathcal{U}_{21} \approx \< \st n_2 | \psi_+\>\<\psi_+ | \st n_1 \> e^{-T(E_{\av}-\frac{1}{2}\Delta E)} + \< \st n_2 | \psi_-\>\<\psi_- | \st n_1 \> e^{-T(E_{\av}+\frac{1}{2}\Delta E)}, \label{eq:u21eigenbasis}
\end{equation}
where the error in the equation above comes from neglecting the higher excited states, which is valid as $T \to \infty$ in the case of spike Hamiltonians because $E'_k \geq E_k$ implies that the density of states for the spike system increases no faster than the density of states of the spikeless system.  After inserting the definitions of $|\psi_+\>$ and $|\psi_-\>$ into \eqref{eq:u21eigenbasis} one finds,
\[
 U_{21} \propto e^{-E_{\av}T}\sinh(T\Delta E),
\]
and this is the result that should be compared with the prediction of $\mathcal{U}_{2 1}$ that comes from the path integral method.  According to \cite{Co88}, when multi-instanton solutions are included the resulting estimate of the full path integral in terms of $S_I$ is
\[
 U_{21} \propto e^{-E_{\av}T}\sinh(T \poly(n) e^{-S_I(n)}),
\]
and so the result \eqref{eq:instantonGapResult} is obtained when both the gap and $e^{-S_I}$ are superpolynomially small so that the hyperbolic sines can be linearized.

\subsection{Agreement with past results}

For the case of cubic cost functions~\cite{FGG02}, 
$E(\theta, \phi)$ is the classical energy given by
\[
 E(\theta, \phi, s) = \left(\frac{n}{2}\right)^3\left(2(1-s)(1-\sin\theta\cos\phi) + sg\left(\frac{1}{2}(1-\cos\theta)\right)\right),
\]
 where
\begin{equation*}
 g(u) = 4qu(1-u)^2 + 4u^2(1-u) + \frac{4}{3}u^3.
\end{equation*}
In this case $C_n = n(n+1)/2$ and $V(\theta, s)$ defined in \cite{FGG02} satisfies
\begin{equation*}
\left(\frac{n}{2}\right)^3 V(\theta, s^*) =  E(\theta, 0, s) = U(\theta, s), \forall \theta, 
\end{equation*}
and substituting this in Eq.~\ref{eq:s0} yields Eqs. (25) and (26) of \cite{FGG02}.
\bibliography{adiabatic}
\bibliographystyle{unsrt}

\end{document}